\documentclass[twoside,twocolumn,9pt]{article}
\usepackage{extsizes}
\usepackage[super,sort&compress,comma]{natbib} 
\usepackage[version=3]{mhchem}
\usepackage[left=1.5cm, right=1.5cm, top=1.785cm, bottom=2.0cm]{geometry}
\usepackage{balance}
\usepackage{mathptmx}
\usepackage{sectsty}
\usepackage{graphicx} 
\usepackage{lastpage}
\usepackage[format=plain,justification=justified,singlelinecheck=false,font={stretch=1.125,small,sf},labelfont=bf,labelsep=space]{caption}
\usepackage{float}
\usepackage{fancyhdr}
\usepackage{fnpos}
\usepackage[english]{babel}
\addto{\captionsenglish}{%
  
}
\usepackage{array}
\usepackage{droidsans}
\usepackage{charter}
\usepackage[T1]{fontenc}
\usepackage[usenames,dvipsnames]{xcolor}
\usepackage{setspace}
\usepackage[compact]{titlesec}
\usepackage{hyperref}

\usepackage{amsmath,amssymb}

\usepackage{epstopdf}

\definecolor{cream}{RGB}{222,217,201}

\begin{document}

\pagestyle{fancy}
\thispagestyle{plain}
\fancypagestyle{plain}{
\renewcommand{\headrulewidth}{0pt}
}

\makeFNbottom
\makeatletter
\renewcommand\LARGE{\@setfontsize\LARGE{15pt}{17}}
\renewcommand\Large{\@setfontsize\Large{12pt}{14}}
\renewcommand\large{\@setfontsize\large{10pt}{12}}
\renewcommand\footnotesize{\@setfontsize\footnotesize{7pt}{10}}
\makeatother

\renewcommand{\thefootnote}{\fnsymbol{footnote}}
\renewcommand\footnoterule{\vspace*{1pt}%
\color{cream}\hrule width 3.5in height 0.4pt \color{black}\vspace*{5pt}} 
\setcounter{secnumdepth}{5}

\makeatletter 
\renewcommand\@biblabel[1]{#1}            
\renewcommand\@makefntext[1]%
{\noindent\makebox[0pt][r]{\@thefnmark\,}#1}
\makeatother 
\renewcommand{\figurename}{\small{Fig.}~}
\sectionfont{\sffamily\Large}
\subsectionfont{\normalsize}
\subsubsectionfont{\bf}
\setstretch{1.125} 
\setlength{\skip\footins}{0.8cm}
\setlength{\footnotesep}{0.25cm}
\setlength{\jot}{10pt}
\titlespacing*{\section}{0pt}{4pt}{4pt}
\titlespacing*{\subsection}{0pt}{15pt}{1pt}

\fancyfoot{}
\fancyfoot[LO,RE]{\vspace{-7.1pt}\includegraphics[height=9pt]{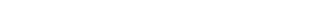}}
\fancyfoot[CO]{\vspace{-7.1pt}\hspace{13.2cm}\includegraphics{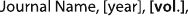}}
\fancyfoot[CE]{\vspace{-7.2pt}\hspace{-14.2cm}\includegraphics{RF}}
\fancyfoot[RO]{\footnotesize{\sffamily{1--\pageref{LastPage} ~\textbar  \hspace{2pt}\thepage}}}
\fancyfoot[LE]{\footnotesize{\sffamily{\thepage~\textbar\hspace{3.45cm} 1--\pageref{LastPage}}}}
\fancyhead{}
\renewcommand{\headrulewidth}{0pt} 
\renewcommand{\footrulewidth}{0pt}
\setlength{\arrayrulewidth}{1pt}
\setlength{\columnsep}{6.5mm}
\setlength\bibsep{1pt}

\makeatletter 
\newlength{\figrulesep} 
\setlength{\figrulesep}{0.5\textfloatsep} 

\newcommand{\topfigrule}{\vspace*{-1pt}%
\noindent{\color{cream}\rule[-\figrulesep]{\columnwidth}{1.5pt}} }

\newcommand{\botfigrule}{\vspace*{-2pt}%
\noindent{\color{cream}\rule[\figrulesep]{\columnwidth}{1.5pt}} }

\newcommand{\dblfigrule}{\vspace*{-1pt}%
\noindent{\color{cream}\rule[-\figrulesep]{\textwidth}{1.5pt}} }

\makeatother

\twocolumn[
  \begin{@twocolumnfalse}
{\includegraphics[height=30pt]{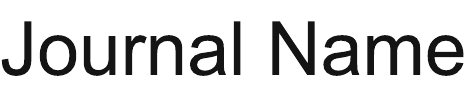}\hfill\raisebox{0pt}[0pt][0pt]{\includegraphics[height=55pt]{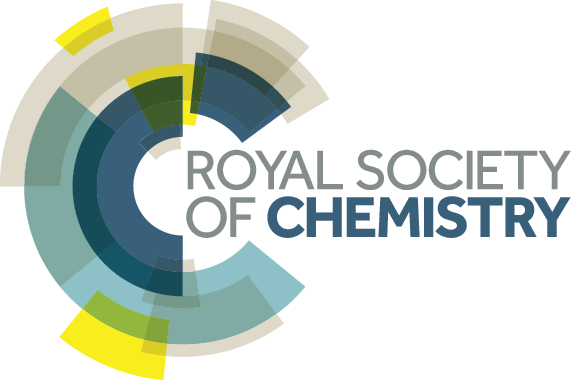}}\\[1ex]
\includegraphics[width=18.5cm]{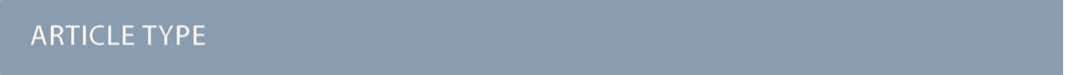}}\par
\vspace{1em}
\sffamily
\begin{tabular}{m{4.5cm} p{13.5cm} }

\includegraphics{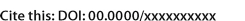} & \noindent\LARGE{\textbf{
MoS$_2$ flake as a van der Waals homostructure: luminescence \mbox{properties~and optical anisotropy} 
}} \\
\vspace{0.3cm} & \vspace{0.3cm} \\

 & \noindent\large{Lyubov V.  Kotova,$^{\ast}$\textit{$^{a}$} Maxim V. Rakhlin,\textit{$^{a}$} Aidar I. Galimov,\textit{$^{a}$} Ilya A. Eliseyev,\textit{$^{a}$} 
 \mbox{Bogdan~R.~Borodin,\textit{$^{a}$} Alexey~V.~Platonov,\textit{$^{a}$}   Demid A. Kirilenko,\textit{$^{a}$} Alexander V. Poshakinskiy,\textit{$^{a}$}} and Tatiana V. Shubina\textit{$^{a}$}} \\

\includegraphics{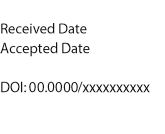} & \noindent\normalsize{We investigated multilayer plates made by exfoliation from a high-quality MoS$_2$ crystal and reveal that they represent a new object -- van der Waals homostructure consisting of a bulk core and a few detached monolayers on its surface. This architecture comprising elements with different electron band structure leads to specific luminescence, when the broad emission band from the core is cut by the absorption peaks of strong exciton resonances in the surface monolayers. The exfoliated flakes exhibit strong optical anisotropy.  We have observed  a conversion of normally incident light polarization to 15~$\%$ in transmission geometry. This background effect is due to fluctuations of the c axis relative to the normal, whereas the pronounced resonance contribution is explained by the polarization anisotropy of excitons localized in the stripes of dissected surface monolayers.
} \\

\end{tabular}

 \end{@twocolumnfalse} \vspace{0.6cm}

 ] 

\renewcommand*\rmdefault{bch}\normalfont\upshape
\rmfamily
\section*{}
\vspace{-1cm}


\footnotetext{\textit{$^{a}$~Ioffe  Institute, St.~Petersburg, 194021, Russia. E-mail:kotova@mail.ioffe.ru}}

\footnotetext{\dag~Electronic Supplementary Information (ESI) available: [details of any supplementary information available should be included here]. See DOI: 00.0000/00000000.}



\section{Introduction}

Transition metal dichalcogenides (TMDC) are among the main materials for modern physics of two-dimensional (2D) semiconductors since 2005, when it was demonstrated that stable free-standing atomic crystals can be obtained by micromechanical cleavage (exfoliation)~\cite{NG}. Among the TMDC family, the molybdenum disulfide (MoS$_2$) occupies a special place as one of the most studied layered compound which in addition is stable under ambient conditions and can be easily synthesized. The 2D MoS$_2$ was investigated for various applications in nanophotonics owing its unique features such as the extremely high exciton binding energy of hundreds of meV and the radiation time of a few picoseconds~\cite{ExcinTMD,Durnev}. Moreover, van der Waals (vdW) heterostructures~\cite{Geim} containing MoS$_2$ monolayers were successfully created~\cite{Vertandinplane,Inter,2DTMDS}. These provide an ideal platform for both fundamental research and new device invention.

Theoretically, it has been predicted that the band gap of bulk MoS$_2$ is indirect being in the near infrared range ($\sim$1.3 eV)~\cite{Mattheiss}, whereas the band gap of a 2D monolayer is direct ($\sim$1.8 eV) and occurs at the high-symmetry point K~\cite{Electronic,Elstruc} of the Brillouin zone. It is important to emphasize that the absorption spectra measured in multilayer MoS$_2$ plates have two strong resonances in the 1.8-2.1 eV range, associated with A and B direct excitons~\cite{Evans,NevEv}. As expected, no features were observed near the indirect band gap. However, an unexpected fact is the lack of data on near-band-edge emission in the bulk MoS$_2$ crystals that might be near 1.3 eV. Only weak PL lines attributed to intercalated halogen atoms and deep donor centers were previously observed between 0.8 and 1.2 eV~\cite{Kulyuk}. The development of the micro-photoluminescence spectroscopy, which provides a much higher excitation power required to detect indirect band-gap radiation, made it possible to obtain experimental confirmation of the transformation of the band structure in the monolayer limit. It showed that a monolayer does exhibit one emission band near 1.9 eV, while stacks of few monolayers (up to 6) demonstrated both direct and indirect PL bands at 1.9 eV and 1.3 eV of much lower intensity~\cite{Mak,Splendiani}. It was suggested that the presence of the disorder in the investigated natural crystals provides the broadening of PL spectra~\cite{polish,AdMat}. Mechanical stress can also lead to polarization splitting of the exciton peaks~\cite{Mitioglu2018}.  Our recent micro-PL studies showed that such two PL bands exist in nanostructures consisting of several tens of monolayers; however, indirect radiation is completely extinguished at low temperatures~\cite{Shubina}. This phenomenon, which requires further study, is probably associated with the temperature balance between fast direct and slow indirect exciton transitions, as well as with the possible contribution of dark exciton states~\cite{Smirnova}. One can assume that modern spectroscopy can shed light on the optical properties of bulk structures with hundreds or thousands of monolayers, which are the parent materials for 2D nanostructures.
Additionally, layered TMDCs naturally possess a giant optical anisotropy in the wide spectral range between the in-plane directions and $c$-axis, which can be useful for on-chip polarization conversion~\cite{natureL}. 

 The widely used exfoliation method inevitably means stressing and bending of a plate that weaken the bonds between the layers, especially between those which are superficial in relation to the rest of the bulk core. A change in the dielectric environment and in the magnitude of the van der Waals gap can lead to the transformation of the local band structure of the superficial layers~\cite{Peelaers}. Besides, the enhanced strain and bending during the mechanical cleavage can cause the formation of oriented extended defects such as ripplocations and  hyperdislocations~\cite{Kushima,Ly}. They separate the layers into domains that is the basic reason of the extremely small size of monolayers fabricated by the exfoliation method (tens of microns only). 

In this work, we have studied samples of micrometer thickness prepared by exfoliation. Measured luminescence spectra with sharp dips at the energies of  A and B exciton resonances, as well as  in-plane optical anisotropy in transmission, show that we are dealing with a new object – \textit{van der Waals homostructures}, consisting of a bulk core and a few detached monolayers on its surface. The appearance of homostructures of one material MoS$_2$ is possible due to the sharp difference between the band structures of the monolayer and the bulk material. The study of the optical properties of such homostructures opens up the possibility of obtaining data on the main characteristics of both a bulk material and a 2D monolayer located on a related semiconductor matrix.

\begin{figure}[b!]
\centering
  \includegraphics[width=1\columnwidth]{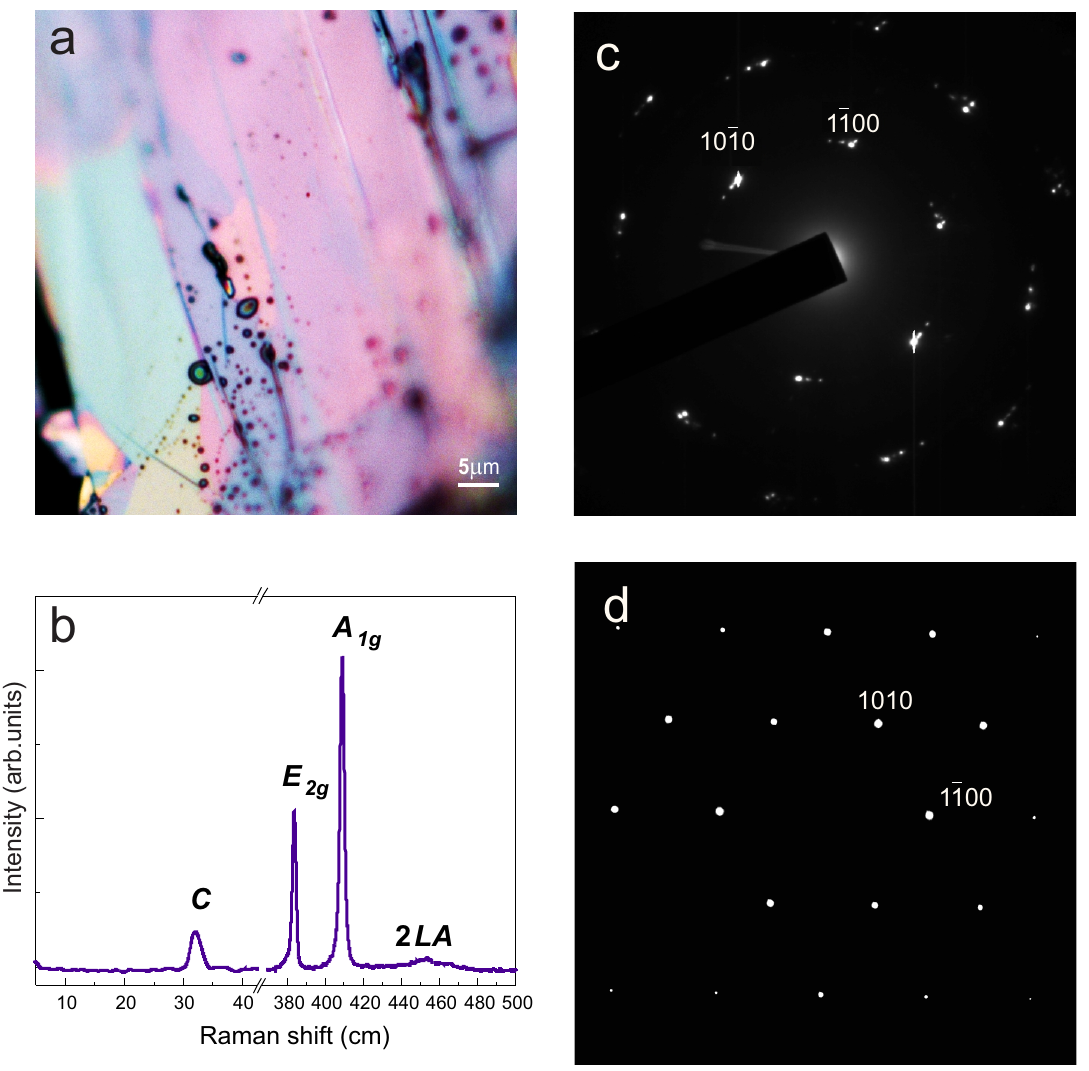}
  \caption{a) Optical microscope image of a sample exfoliated from the bulk MoS$_2$ crystal using adhesive-tape method (the bubbles are likely traces of the adhesive tape).
  b) A Raman spectrum measured at 300~K, which is typical for the entire surface of the sample. 
  c), d) Diffraction patterns of (c) the mechanically cleaved flake, and (d) an uncleaved thin flake synthesized by the chemical transport reaction method. }
  \label{figura1}
\end{figure}

\section{Samples}

The vdW structures under study were fabricated from a commercially available high-quality  MoS$_2$ bulk crystal (HQ-graphene production). For sample preparation, both mechanical microcleavage and a so-called Scotch-tape method \cite{tape} were used. An optical microscope image of one of such flakes fabricated  using adhesive tape technology is presented in Fig.~\ref{figura1}~a and also more details in Supplement Fig.~S1. The sample surface consists of domains separated by narrow gaps; the color inside the gaps looks the same as in the underlying core.

The thickness of the samples prepared for optical studies was determined by the interference in the transmission spectrum and equals 1~$\mu$m. Structural quality of the samples is typical for high-quality MoS$_2$, as confirmed by micro-Raman studies, namely: by the observed Raman peak positions and their narrow width (see Fig.~\ref{figura1}~b). The Raman signal is identical for all detection areas.  Regardless of the signal registration area, the obtained Raman spectra did not display any differences between each other. All features visible in the spectrum (Fig.~\ref{figura1}~b), such as the $C$ ($31 \, cm^{-1}$), $E_{2g}$ ($383\, cm^{-1}$) and $A_{1g}$ ($409\, cm^{-1}$) lines are characteristic of high-quality MoS$_2$\cite{Raman, big,Ram}. Frequency difference between the $A_{1g}$ and $E_{2g}$ Raman lines ($26 \, cm^{-1}$) corresponds to the values reported for bulk MoS$_2$~\cite{Ram1}. Presence of an interlayer $C$ mode with a frequency of $31 \, cm^{-1}$ together with the absence of layer-breathing modes in the $5-30 \, cm^{-1}$ allows us to characterize the material as a bulk MoS$_2$ crystal~\cite{Ram2}. Note that the Raman signal is collected from the entire thickness of the sample, which masks the possible contribution of surface monolayers (see Supplement Fig.~S2).


\begin{figure}[t!]
\centering
  \includegraphics[width=0.95\columnwidth]{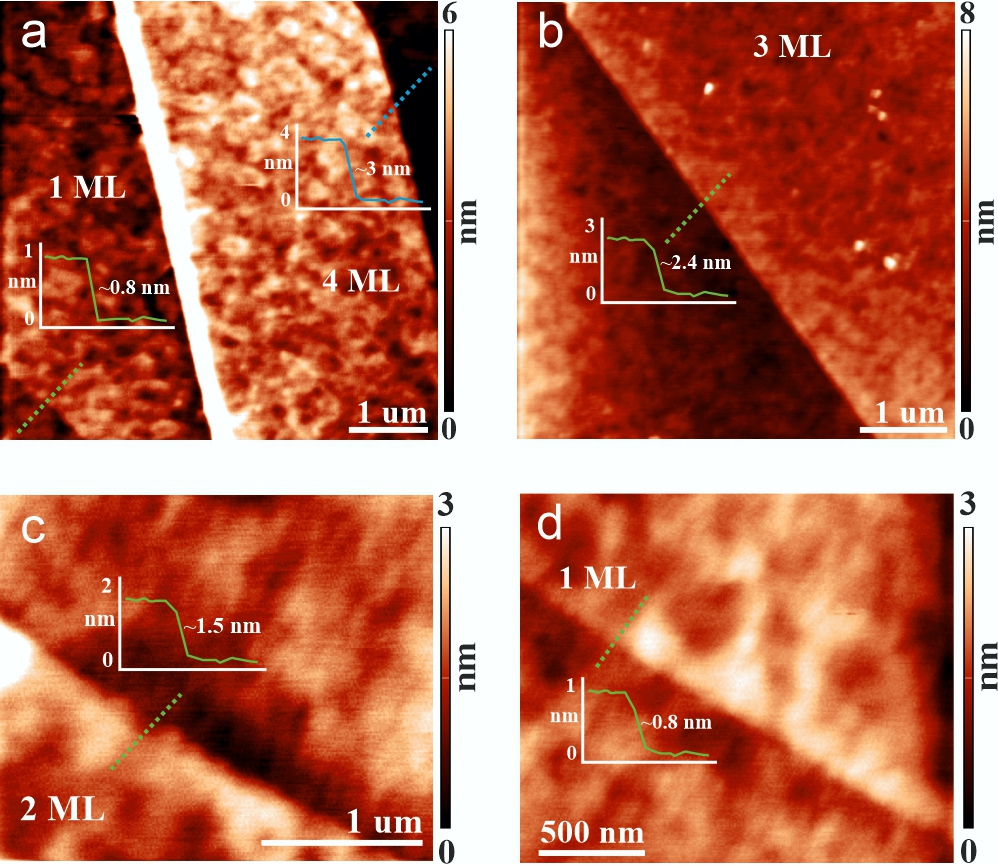}
  \caption{AFM images of the flake's surface obtained after last step of the exfoliation procedure. Panels (a)-(d) correspond to the  same points of the flake as in Supplement Fig.~S4  which differ by the number of few detached monolayers on the surface.}
  \label{fgr:figura_AFM}
\end{figure}

To determine how the delamination process affects the crystal structure, we conducted transmission electron microscopy (TEM) studies. The results of the TEM studies of samples fabricated from a bulk MoS$_2$ crystal by mechanical cleavage are presented in Fig.~\ref{figura1}~c,d. 
Investigation of the fragments cleaved along the (0001) plane suggests that 
there is a spatial misorientation  up to 10 degrees between different  regions
in the sample. This misorientation leads to the multiplication of the diffraction spots in  Fig.~\ref{figura1}~c. This feature is absent in the diffractogram of an intact flake synthesized by the chemical transport reaction method, as described in~\cite{Shubina}. More detailed images of TEM investigation are in Supplement Fig.~S3,4.

To study the topography and the number of layers (thickness) of the flakes cleaved, atomic force microscopy (AFM) was used. Results of the investigation are presented in Fig.~\ref{fgr:figura_AFM}.
As can be seen, regions of different thicknesses (from 0.8 to 3 nm) can be identified on the surface. It is well known that the averaged thickness of a monolayer may slightly differ depending on the substrate and interlayer conditions, and typically equals 0.65-0.9 nm~\cite{Lanzillo}. Here, we found the minimum step of 0.8 nm that can be referred to monolayer. We report that the surface of the studied structure is covered from 1 to 5 detached monolayers in different regions.

In general, all this results show that sample preparation provides suitable conditions not only for the formation of defects, but also for the separation of atomic layers. This trend should be more pronounced for superficial atomically-thin layers. As a result, the cleaved flakes can be viewed as homostructures consisting of a bulk MoS$_2$ core and a few monolayers at its surface or somewhere within the volume, that cannot be excluded. To find additional facts confirming such architecture, we performed optical studies as described below.


\section{Emission properties}
\begin{figure}[h]
  \includegraphics[width=1\columnwidth]{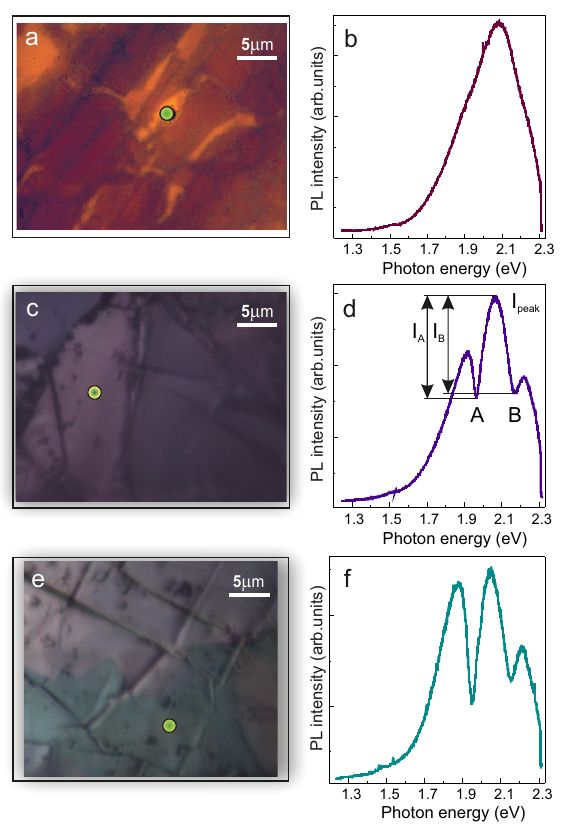}
  \caption{a,c,e) The surface image of the MoS$_2$ homostructure.  The area where $\mu$-PL was excited is shown by green dot. b,d,f) $\mu$-PL intensity spectra. The position and amplitude of the dips corresponding to light absorption by A and B excitons in surface monolayers are indicated. T=10K.}
  \label{figura3}
\end{figure}

 
 The photoluminescence of MoS$_2$, which thickness varies from monolayer to bulk, has been already well studied~\cite{Confocal,Mak,Eda,nanomat,Tightly,ExcinTMD,GlazovM}.  
Here we demonstrate that the photoluminescence of the MoS$_2$ vdW homostructures, which arise naturally during exfoliation, are rather distinct.  

\subsection{Micro-photoluminescence spectra}

 \begin{figure*}[h!]
\centering
 \includegraphics[width=2.05\columnwidth]{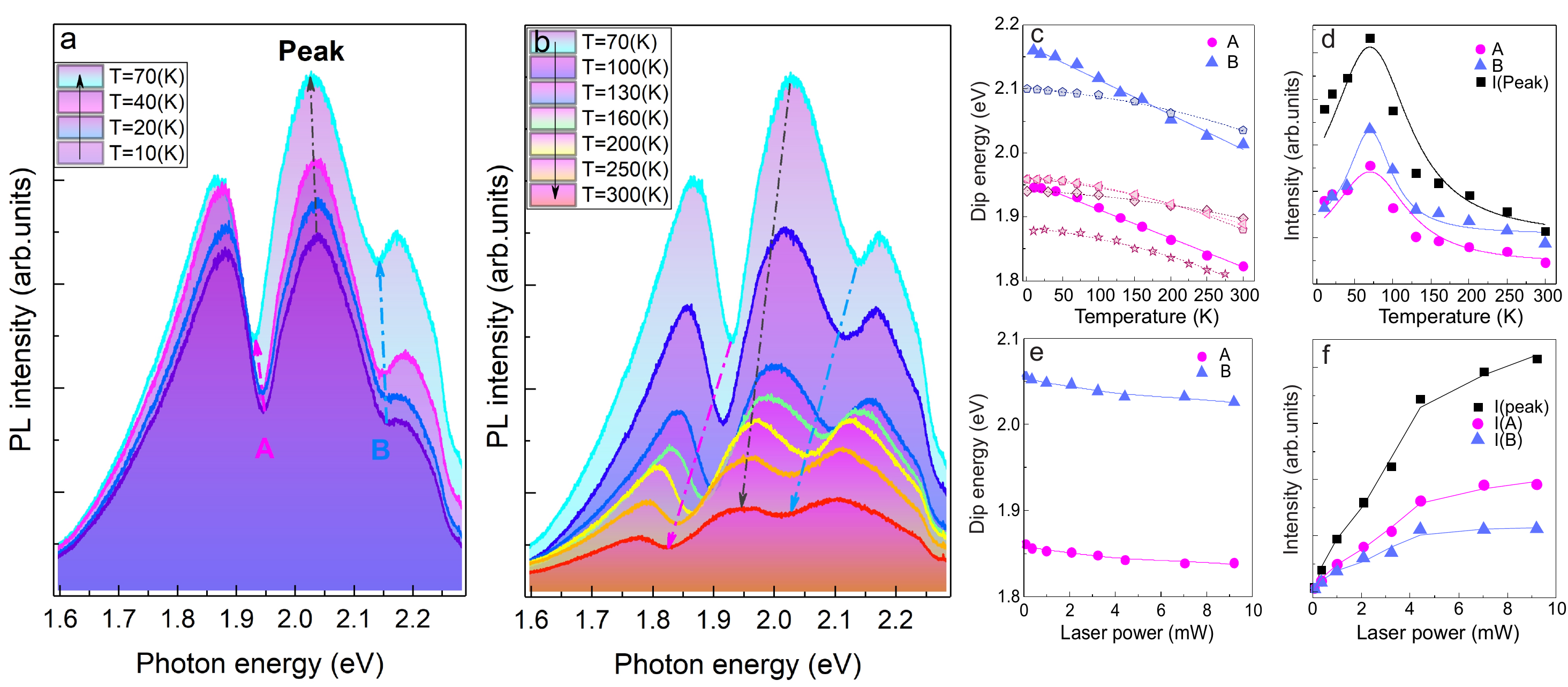}
 \caption{a,b) $\mu$-PL intensities at different temperature. A and B excitons positions are indicated on the plot.
 c,f) Dependence of the position of A and B excitons 
 on temperature and laser pump power. The unpainted characters represent the literature data on excitons in monolayers(triangle~\cite{Liu}, rhombus~\cite{ZHANGPL}, pentagon~\cite{Hguen}, hexagon~\cite{Arora}), 
asterisks represent the PL peak energy in the synthesized uncleaved thin flake. d,e) The $\mu$-PL intensities dependencies on the temperature and laser pump power. Laser pump power dependency was obtained at room temperature.}
 \label{fgr:figura34}
 \end{figure*}
 
Micro-photoluminescence 
spectra obtained from different points on the surface are shown in Fig.~\ref{figura3}~a-f. The spectra are of three distinct types.  Figures~\ref{figura3}~a,b correspond to the case when the signal is collected from the bulk MoS$_2$ core of the sample. The broad peak in the luminescence is explained by the near-band-edge transitions with some  contribution of excitonic emission. On the contrary, the signal collected from the areas shown in Fig.~\ref{figura3}~c,e features two narrow dips atop of the broad peak corresponding to bulk luminescence. In this flat area, the bulk core of the homostructure can be covered by a few detached MoS$_2$ monolayers. The dips are explained by the light absorption at frequencies close to exciton resonances in the surface monolayers, namely, A exciton at 1.94~eV and B exciton at 2.15~eV (at T=10 K).  
We assume that the areas described in Fig.~\ref{figura3}~c,d and Fig.~\ref{figura3}~e,f differ in the number of monolayers covering the bulk core, which leads to different amplitudes of the dips (see comparison in Supplement Figure S6). 
The position of exciton resonance is almost independent from the number of layers due to the interplay of the band gap change and exciton binding energy variation~\cite{PhysRevB.90.205422}. 
The exciton absorption for a monolayer can be estimated as $A_1=2\Gamma_0/\Gamma_x$, where $\Gamma_0$ is the radiative exciton decay and $\Gamma_x$ is the non-radiative broadening of exciton resonance, and $\Gamma_0 \ll \Gamma_x$ is assumed. For A exciton, using $\Gamma_0 = 1.4\,$~meV~\cite{Palummo2015} and $\Gamma = 30\,$~meV, as extracted from the HWHM of the exciton dip (see Supplement Figure S7), we estimate that $N \sim 5$ surface monolayers are required to achieve the PL attenuation $(1-A_1)^N \approx 0.6$ observed at exciton resonance. It is in excellent agreement with the experimental results of AFM measurements of detached monolayers on the  investigated vdW surface.   
Due to the higher energy of the conduction band minimum in MoS$_2$ monolayer as compared to bulk material~\cite{Molina},  the monolayers on the top of the homostructure are electron depleted and their luminescence is quenched, i.e. intensity is suppressed like it occurs for barrier emission in a heterostructure. In the measured PL spectra, it is masked by the much stronger bulk emission.

In addition to the direct exciton features, the signal collected from some areas features also peak at the indirect exciton frequency $\sim$1.3~eV, see Supplementary Figure S5~a,b.  The dominance of the direct-gap and 
transitions in the spectra of multilayer structures at low temperature was previously observed in Ref.~\cite{Shubina} and explained by the fast radiative recombination of the direct excitons as compared to slow  energy relaxation and phonon-assisted recombination of indirect excitons~\cite{Smirnova}. In addition, at a few points purely excitonic radiation dominates, as shown in the Supplementary Fig. S5~c,d.



 \begin{figure*}[h]
\centering
 \includegraphics[width=2.05\columnwidth]{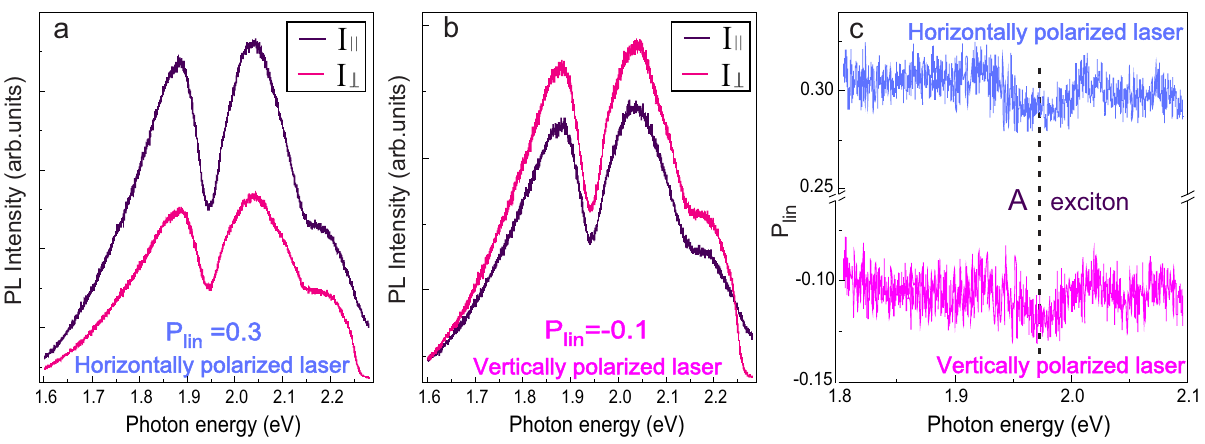}
 \caption{
  {
 a,b) The $\mu$-PL spectra measured in linear polarizations parallel ($I_\parallel$) and perpendicular ($I_\perp$) to the polarization of the laser. Panels (a) and (b) correspond to the two orthogonal polarizations of the laser. Panel (c) shown the degree of linear polarization $P_{\rm lin}$ calculated after Eq.~\eqref{S}. 
 The measurements were done at $T=10$\,K. 
 } }
 \label{fgr:PolPL}
 \end{figure*}

 
\subsection{Dependence on the temperature and excitation power} 
 
We also measured the evolution of the $\mu$-PL spectra with temperature, Fig.~\ref{fgr:figura34}~a,b and pump power, Supplementary Fig. S8. Figure~\ref{fgr:figura34}~c-f shows dependence of the positions and intensities of the broad peak (bulk luminescence) and two dips (exciton absorption).
The positions of the dips decrease with temperature, see Fig.~\ref{fgr:figura34}~c, that generally agrees with the variation of the band gap. Energetically decreasing dependencies were observed in multiple previous optical absorption and PL studies, see unpainted characters in Fig.~\ref{fgr:figura34}~c and the Supplementary Tables~I and II. However, the dependencies do not match with the slope with ours, since the vdW homostructure is different from both synthesized flake, that is, the ideal bulk, and a stack of monolayers on different dielectric substrates.
Note that the broad peak position can be determined with limitted accuracy  due to its complex nonsymmetric shape additionally modified by the exciton dips. 
The intensities of the peak and the dips increase at small temperatures, then reach maximum at around 70\,K and decrease at higher temperatures, see Fig.~\ref{fgr:figura34}~d.  Such behavior could indicate that the transition with the lowest energy is dark in the bulk MoS$_2$ as was asuumed in Refs.~\cite{Shubina,Malic}.  In this case, the bright optical transition with higher energy starts emitting only with an increase of the temperature.
The release of carrier trapped by defects could also contribute to the increase of PL at low temperatures.
%
With an increase of the laser power the positions of the dips remain almost constant, see Fig.~\ref{fgr:figura34}~e. A very small shift occurs probably due to the heating by a laser beam. The intensities of the peak and the dips grow linearly and then saturate, see Fig.~\ref{fgr:figura34}~f.  Their ratio remains almost constant, as shown in Supplement Fig. S9.


{
\subsection{Polarization of the emitted light}

Finally, we studied the PL polarization. The excitation was performed by the light with a certain linear polarization, while the PL spectra were recorded in six different polarizations:  polarized along ($I_{\parallel}$) and perpendicular ($I_{\perp}$) to the incident light polarization, polarized along the axes rotated by $\pm 45^\circ$ with respect to the incident light polarization ($I_{1,2}$), and  polarized circularly ($I_{\sigma_\pm}$).
From these spectra, the full set of the Stokes parameters can calculated according to
\begin{equation}
\label{S}
		P_{\rm lin}={I_{\parallel}-I_{\perp} \over I_{\parallel}+I_{\perp}}	
			\quad
		\tilde{P}_{\rm lin}={{I}_1-{I}_2 \over {I}_1+{I}_2},
		\quad 
		P_{\rm circ}={I_{\sigma_+}-I_{\sigma_-} \over I_{\sigma_+}+I_{\sigma_-}}.
\end{equation}
Figure~\ref{fgr:PolPL}a,b shows the $I_{\parallel,\perp}$ spectra in the two cases corresponding to the two orthogonal polarizations of the laser. The degree of emitted light linear polarization $P_{\rm lin}$ calculated according to Eq.~\eqref{S}  is almost constant in the wide spectral range, see Fig.~\ref{fgr:PolPL}c.

The linear polarization of PL could originate from the intervalley coherence, but in that case the polarization degree would be independent from the incident laser polarization orientation~\cite{Cadiz,Jones}. However in our case,   $P_{\rm lin}$ appears to have different values and opposite signs for the two laser polarizations. This indicates that the structure possesses optical anisotropy which dictates for the emitted light a certain polarization direction  that is distinct from the polarization of the laser.
 We have also measured the linear polarization degree in the axes rotated by $\pm 45^\circ$ with respect to the incident light polarization, $\tilde{P}_{\rm lin}$, which demonstrated the behaviour analogues to $P_{\rm lin}$, and the circular polarization degree, $\tilde{P}_{\rm circ}$, which appeared to be zero up to the experimental noise (not shown). The latter fact indicates that the structure is not chiral, thus the type of optical anisotropy is identified as linear birefringence.   
A careful inspection of the $P_{\rm lin}$ spectral dependence in Fig.~\ref{fgr:PolPL}c
reveals the presence of the small peak  marked by a vertical dashed line.  The peak position  matches perfectly with the position of the exciton peak in the reflectance spectra, see Supplementary Fig.~S10. Exciton features are visible in the spectra up to room temperature.  The peak amplitude is about an order of amplitude smaller than the constant background.
Below we study the optical anisotropy of the sample in detail and demonstrate that it indeed has two contributions of different strength, spectral dependence, and symmetry, thus proving the proposed model of a vdW homostructure that consists of the surface and the bulk region.


}

\section{Optical anisotropy}

\begin{figure}[t]
 \centering
 \includegraphics[width=0.98\columnwidth]{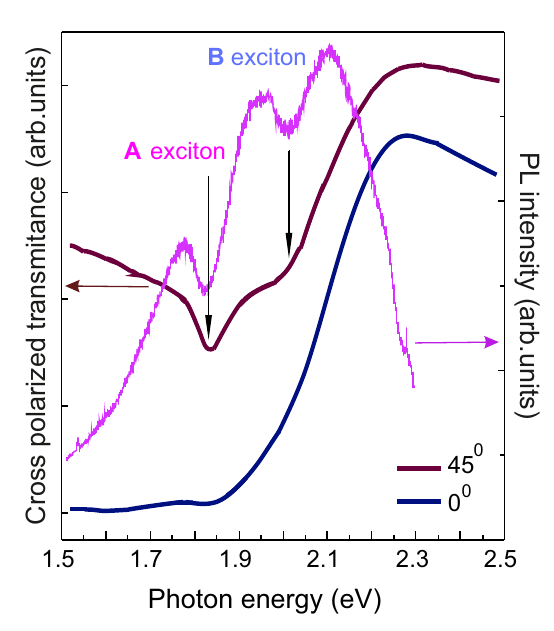}
 \caption{The transmission spectra  of the  MoS$_2$ van der Waals homostructure in cross polarization for two angles of sample rotation measured at room temperature, shifted for clarity and shown together with a PL spectrum.}
 \label{fgr:figura3a}
 \end{figure}
 
\begin{figure*}[t]
\centering
\includegraphics{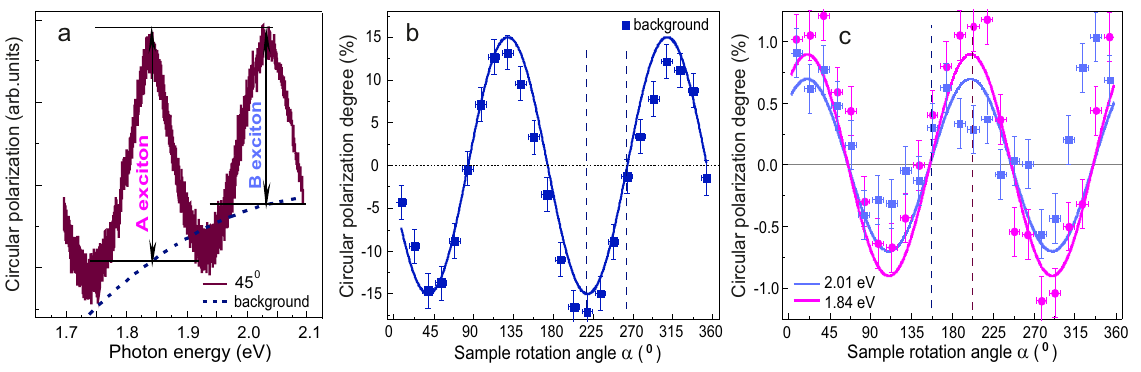}
\caption{a) The spectra of the transmitted light circular polarization degree for the  MoS$_2$ vdW homostructure measured at orientation corresponding to the the maximal signal. 
The background signal and the amplitudes of the peaks related to A and B excitons are indicated. b) Sample orientation dependence of the P$_{circ}$  for background signal valid in a wide spectral range. c) Sample orientation dependencies of A and B exciton peaks' amplitudes. All curves are obtained at room temperature.}
 \label{fgr:figura3}
 \end{figure*}
 
\begin{figure}[h!]
 \centering
 \includegraphics[width=0.9\columnwidth]{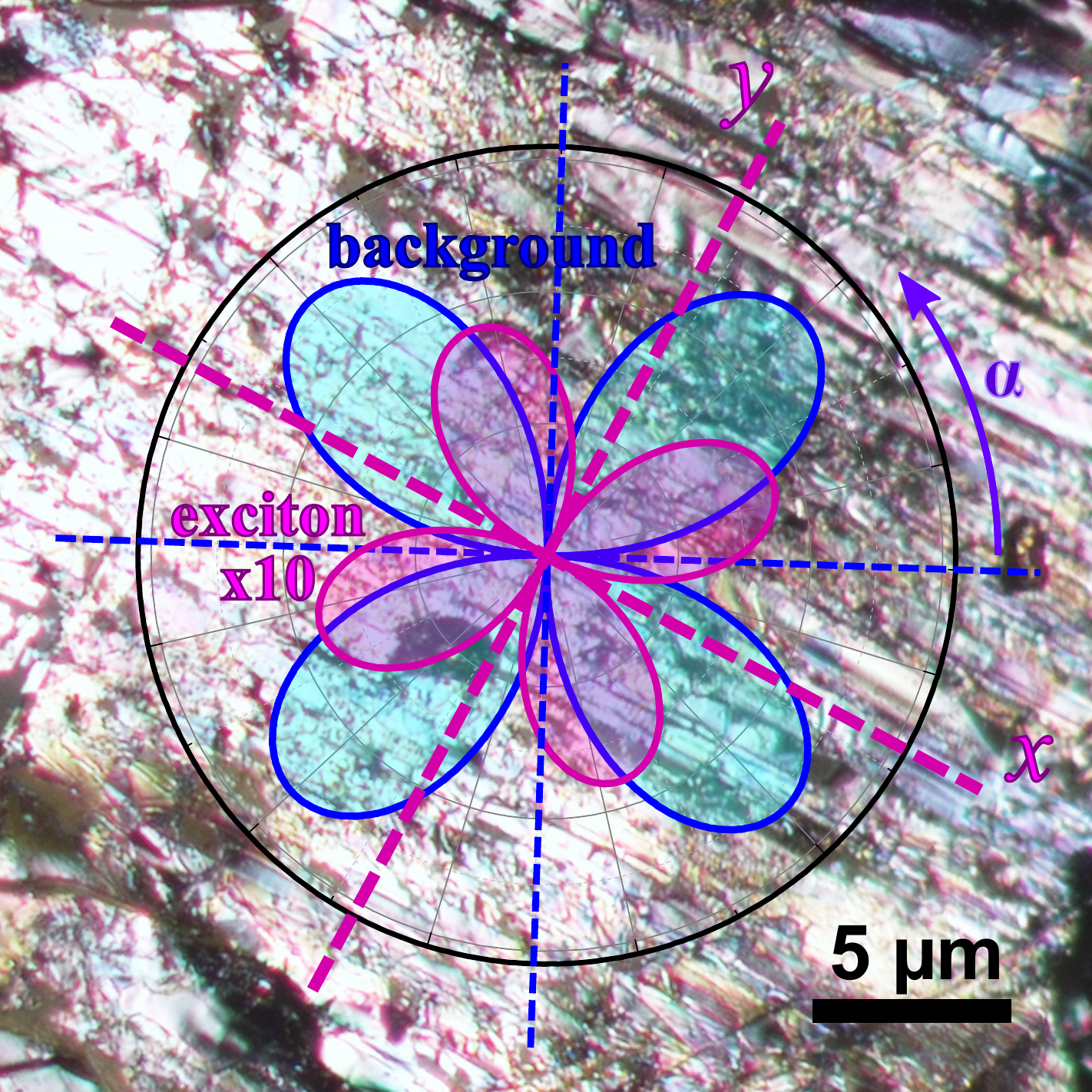}
 \caption{MoS$_2$ homostructure's surface image with bright-field microscopy obtained by Nikon industrial microscope ECLIPSE LV150. Over it there are two angular dependencies of circular polarization for transmitted light. Navy blue feature corresponds to the background anisotropy stemming from the bulk core. Magenta curve corresponds to the anisotropy in the vicinity of the exciton resonance (the signal value is multiplied by 10 to facilitate comparison of directions).}
 \label{fgr:figura4}
 \end{figure}
 

To quantify the optical anisotropy, we study the coherent light transmission though the sample. The structure was excited by linear polarized light incident along the normal and the full set of transmitted light Stokes parameters~Eq.~\eqref{S} was measured.  Such technique allowed us to achieve a higher signal-to-noise ratio for the measured Stokes parameters as compared to 
PL measurements discussed above. 
Similar approach has been used 
recently to study the optical activity of QWs~\cite{OptActiv, LVK,Interf}.

\subsection{Polarization conversion at transmission}

The spectra of the transmitted light component cross polarized to the incident light $I_{\perp}$ measured at room temperature are shown in Fig.~\ref{fgr:figura3a} for two orientations of incident light polarization which correspond to minimal (blue curve) and maximal (red curve) signals. 
The strong dependence of $I_{\perp}$ on the incident light polarization orientation is present in a wide spectral range and indicates the in-plane optical anisotropy of the structure, i.e., the difference of the dielectric permittivities for the two in-plane directions. Such optical anisotropy can arise due to in-plane stress or shear deformation of the bulk core. Here, it is most likely caused by the anisotropic fluctuations of the flakes' $c$-axes directions, revealed in testing X-ray measurements (the studied sample is too small to register XRD scans).     
The transmitted light linear polarization degree in the axes rotated by $\pm 45^\circ$, $\tilde{P}_{\rm lin}$, is found to be very small (less than 2\%, see Supplementary Figure S11), which indicates that linear dichroism does not exist in the structure, i.e, the imaginary part of the dielectric permittivity is isotropic in the plane of the structure.

On top of the smooth background, the cross-polarized transmittance spectra (Fig.~\ref{fgr:figura3a}) feature peculiarities at the frequencies of A and B excitons. Their positions agree with the previous investigations of MoS$_2$ thin crystals~\cite{Evans,Wilson,Frindt,13}. 
To further investigate the optical properties associated with the exciton transitions in the vdW homostructure, we have measured the transmitted light circular polarization degree  $P_{\rm circ}$ in the corresponding frequency range, see Fig.~\ref{fgr:figura3}. Similarly to cross-polarized transmittance, the spectrum of the circular polarization degree, Fig.~\ref{fgr:figura3}a, consists of a smooth background with two exciton peaks on top of it. Dependence of the background value and of the exciton peak amplitudes on the sample rotation angle $\alpha$ is shown in Fig.~\ref{fgr:figura3}~b,c. The  measured data are well described by the sine function
\begin{align}
P_\text{circ}(\alpha) = \frac{\omega d}{2c}\, \text{Re}(\Delta n)\, \sin 2(\alpha-\alpha_0)\,,
\end{align}
where $d$ is the sample thickness, $\Delta n$ is the difference between  the refractive indices of the two in-plane optical axes, $\alpha_0$ is the direction of one optical axis, and the direction of the other is $\alpha_0+90^\circ$.  When the incident light is  polarized along one of the optical axes,  $P_{\rm circ}$ vanishes, while for directions rotated by $\pm 45^\circ$ with respect to them $P_{\rm circ}$ is maximal. 
The absolute value of the circular polarization degree allows us to quantify the strength of the in-plane optical anisotropy. The background contribution is determined by the the bulk of the structure, so we use $d=1\,\mu$m and obtain $\Delta n^\text{(bg)} = 1.2 \cdot 10^{-2}$. The excitonic contribution is determined only by the surface monolayers of width $d_x$ and is an order of magnitude weaker. The fit gives 
$\Delta n^\text{(x,A)} \, d_x = 9 \cdot 10^{-4}\,\mu$m and $\Delta n^\text{(x,B)} \, d_x = 7 \cdot 10^{-4}\,\mu$m for A and B excitons respectively. Using $d_x \sim 10\,~\text{nm} $, we estimate $\Delta n^\text{(x)} \sim 0.1$.

Interestingly, the optical axis direction $\alpha_0$ extracted from the angular dependence of the exciton peak amplitude  and that of the background are different, cf. the phases of the cosine in Fig.~\ref{fgr:figura3}~b,c. As discussed in Sec.~3.1, the excitonic features are determined by the few detached monolayers on the surface of  the homostructure. Therefore, we conclude that the optical axes of the surface are different from those of the bulk by $\alpha_0^\text{(x)} - \alpha_0^\text{(bg)} = 27^\circ$. To prove that, we take the picture of the surface using bright-field microscope, see Fig.~\ref{fgr:figura4}. The image reveals that linear structural defects aligned in a certain direction are present on the surface~\cite{Bailey}. The presence of such defects is probably  associated with the production of the sample using the adhesive tape method. We superpose the angular dependencies of the background and exciton contributions to the surface image, see the navy blue and magenta curves which are drawn after the corresponding curves in Fig.~\ref{fgr:figura3}~b,c. The orientation of the exciton contribution  turns out to be consistent  with the direction of the structural defects: the conversion to the circular polarization vanishes when the incident light is polarized along or perpendicular to the linear defects, see the optical axes indicated by the dashed magenta lines, and reaches maximum when the incident light polarization is rotated by  $\pm 45^\circ$ with respect to them.

The orientation of the linear defects, which are  the fractures on the surface, most probably coincides with one of the the zigzag  directions $\langle 1\bar100 \rangle$, since cutting material along this direction requires breaking the smallest number of chemical bounds. Then, the axes of the background optical anisotropy, which are rotated with respect to those of the exciton contribution by the angle very close to 30$^\circ$ (the deviation by a few degrees is probably due to the depolarization effect that alters the background signal as it passes through the linear defects on the surface~\cite{Hu2018}), are close to the crystallographic directions  $\langle 11\bar20 \rangle$ and $\langle 1\bar100 \rangle$. The presence of distortions along such directions was revealed in the diffraction patterns of Supplement Fig. S2,S3.  

\subsection{The origin of the optical anisotropy}

Finally, we discuss the microscopic mechanisms for the observed optical anisotropy. The background contribution can be explained by the dispersion of the $c$-axis direction in the bulk of the structure. 
Due to the birefringence of the bulk MoS$_2$\cite{natureL}, a tilt of the $c$-axis in a certain direction from the structure normal to a small angle $\theta$, leads to in-plane optical anisotropy
\begin{align}
    \Delta n^\text{(bg)} = \frac{\sqrt{\epsilon_\perp}}{2}\left( 1 - \frac{\epsilon_\perp}{\epsilon_\parallel}\right) \theta^2 \,,
\end{align}
where $\varepsilon_{\parallel,\perp}$ are the dielectric constants of the bulk MoS$_2$ along and perpendicular to the $c$-axis. Using $\varepsilon_{\parallel} = 6.5$, $\varepsilon_{\perp}=15.5$\cite{Evans,Laturia2018} we estimate the mean square angle of $c$-axis tilt  required to obtain the observed value of $\Delta n^\text{(bg)}$ as $\sqrt{\langle \theta^2 \rangle} \sim 2.5^\circ$.

To explain the exciton contribution to optical anisotropy, we suppose that the excitons in the monolayers on the structure surface get localized on linear defects, being able to move freely along them while confined in the perpendicular direction. The lateral confinement wave vector $K\sim \pi/l$, where $l \lesssim 1\,\mu$m is the width of the linear defect, is comparable to the light wave vector. As a  result, a significant longitudinal-transverse splitting between the frequencies of excitons polarized along and perpendicular to the linear defects arises~\cite{Gupalov1998,Glazov2007}. The splitting value for the excitons inside the light cone depends strongly on the dielectric environment and is limited by the exciton radiative decay rate,\cite{Prazdnichnykh2020} $\Delta\omega_x \lesssim \Gamma_0 $.
Due to this splitting,  the refractive index  acquires an in-plane anisotropy
\begin{align}
\Delta n^\text{(x)} =  -\frac{\sqrt{\varepsilon_\perp} \omega_{\rm LT} \Delta\omega_x}{2(\omega-\omega_x+i\Gamma_x)^2} \,,
\end{align}
where $\Gamma_x$ is the exciton width and $\omega_{\rm LT}$ is  the exciton-induced longitudinal-transverse splitting in the surface layer of the homostructure covered by a few MoS$_2$ monolayers. The latter can be calculated as $\omega_{\rm LT} = 2c\Gamma_0/(\omega a)$,\cite{Kazanov2020} where $\Gamma_0$ is the radiative decay rate of a single monolayer and $a$ is the the distance between the monolayers.  
Using $\Gamma_x = 50\,$meV, as determined from the spectrum in Fig.~\ref{fgr:figura3}(a) and the surface monolayer thickness $d_x \sim 5 a$, we estimate that to obtain $\Delta n^\text{(x)} d_x \sim 10^{-3}\,\mu$m observed in the experiment, the exciton energy splitting should be $\Delta\omega_x \sim 0.8\,$meV, which is safely below the theoretical upper limit for the in-plane longitudinal-transverse splitting $\sim \Gamma_0 = 1.4$\,meV.  

Depolarization of the electric field inside the linear defects could result in a difference of the exciton oscillator strength along and perpendicular to the defects and  result in refractive index anisotropy of the form $\Delta n^\text{(x)} = \sqrt{\varepsilon_\perp} \Delta\omega_{\rm LT} /(\omega-\omega_x+i\Gamma_x)$. However, the corresponding contribution to circular polarization degree, determined by $\text{Re\,} \Delta n^\text{(x)}$, would have a form of the Lorentz function derivative, vanishing at the exciton resonance frequency, in contradiction to a peak at exciton frequency observed in experiment.




\section{Conclusion}

To conclude, the van der Waals homostructure consisting of a bulk core and a few detached monolayers on its surface was fabricated and investigated. This architecture, consisting of elements with different structures of electronic bands, leads to specific luminescence spectrum: the broad emission band of the bulk core is cut off by the absorption peaks of strong exciton resonances in the surface monolayers. 
 We have studied the optical transmission of MoS$_2$ vdW structures and revealed their polarization anisotropy leading to polarization conversion. The effect consists of two contributions: (i)  background anisotropy due to linear birefringence of the bulk crystal and the fluctuations of the $c$-axis direction and (ii) longitudinal-transverse splitting of the exciton energies due to their localization by the structural defects on the sample surface. The bulk contribution leads up to 15\% linear-to-circular polarization conversion per micron of sample thickness. The exction contribution is one order of magnitude smaller and its angular dependence is consistent with the direction of  linear structural defects on the sample surface.
 
In our experiments, the formation of MoS$_2$ homostructures is the result of an exfoliation procedure that peels off the surface monolayers. This is accompanied by the formation of linear defects that dissect the detached surface monolayers into domains only tens of microns wide. It should be emphasized that these effects set a natural limit on the maximum size of monolayer crystals that can be obtained by the Scotch tape method for chip technology. On the other hand, this opens the way to the study of a unique homostructure, whose components have the same chemical composition, but have different electronic band structures of 3D and 2D crystals. This can shed light on the main characteristics of both bulk crystals and monolayers, as well as outline possible applications of such homostructures. Homostructures with more sophisticated optimized design can be used for efficient light manipulation at nanoscale including lasers, light-emitting diodes, and solar cells. The emission properties of the considered homostructures can be further improved by annealing to eliminate defects in bulk or treatment by organic compound of the detached monolayers on the surface\cite{Amani1065}. 

\section{Methods}

\textbf{Sample preparation}.
Mechanical microcleavage and a so-called Scotch-tape method were used for sample preparation. The most common methods of production of TMDC nanomaterials are techniques involving solution-phase exfoliation or mechanical microcleavage, or chemical vapour deposition\cite{D0MA00697A}.

\textbf{Micro-photoluminescence ($\mu$-PL)} setup was used for luminescence properties
investigation of MoS$_2$ van der Waals homostructures. The sample was mounted in a He-flow cryostat ST-500-Attocube with a XYZ piezo-driver inside, which allowed us to optimize and precisely maintain the positioning of a chosen place on the sample with respect to a laser spot during a long time (few hours). A micro-cryostat allowed the measurements at different temperatures. Experimental data were obtained at 10~K. Non-resonant optical excitation of a cw Nd:YAG laser (532 nm) was used for the $\mu$-PL measurements. Incident radiation was focused in 2-3 $\mu$m spot on the sample by an apochromatic objective lens with NA of 0.42. The power density was $\sim$ 6 W/cm$^{-2}$. The collected emission was dispersed by a gap 0.5~m monochromator with a 600/mm grating for detection $\mu$-PL spectra at selected wavelengths. 
In addition, the Raman lines were recorded simultaneously while scanning the PL bands. The PL spectra were normalized to the intensity of the Raman line to exclude  random variations.

\textbf{Raman}. 
Another setup was dedicated for micro-Raman~($\mu$R) and PL spectroscopy at room and liquid nitrogen temperatures, respectively.  This setup consisted of a Horiba Jobin-Yvon~T64000 spectrometer equipped with a confocal microscope and a silicon CCD detector. Use of an objective lens with a 100x magnification and a NA of 0.9 together with a 1800 gr/mm grating allowed us to obtain Raman signal at room temperature with high spectral resolution from an area of $\sim$1~$\mu$m in diameter. The temperature controlled microscope stage Linkam~THMS600, long working distance lens (50x magnification, 0.5 NA), and a 600 gr/mm grating allowed us to carry out temperature dependent PL measurements with a spatial resolution of 2-3 microns. For excitation of the Raman also a 532~nm (2.33~eV) line of a Nd:YAG laser was used.

\textbf{TEM}.
To determine how the delamination process affects the crystal structure, we conducted transmission electron microscopy studies using a Jeol JEM-2100F microscope. The results of the TEM studies of samples fabricated from a bulk MoS$_2$ crystal by mechanical cleavage~\cite{TEM} are presented in Fig.~\ref{figura1}~c,d and in Supplement Fig. S2,S3.

Atomic force microscopy (\textbf{AFM}) was used to study topography and the number of layers (thickness) of the cleaved flakes on the surface. AFM investigation was performed on an Ntegra Aura (NT-MDT, Russia) scanning probe microscope using Si probes (HANC, TipsNano) with resonant frequency $f_0 \approx 140 $ kHz, spring constant $ k \approx 3.5 $ N  m$^{-1}$ , and tip curvature radius < 10 nm. Experimental data are presented in Fig.~\ref{fgr:figura_AFM}.

\textbf{Transmission}.
To study the transmission of the MoS$_2$ van der Waals homorostructures, a setup was used that also made it possible to measure the polarization properties. Polarization control was achieved by means of a half- and quarterwave  plates and a Glan-Taylor prism installed in the detection channel. Different polarization configurations were used. The incident light was linearly polarized. Spectral dependencies of the transmitted light intensity $I(\omega)$ were measured. Namely, two circular intensities $I_{\sigma_\pm}$, two linear ones $I_{\perp,\parallel}$, and two linear components in the axes rotated by $\pm45^\circ$ relative to the plane of incidence,  $I_{1,2}$. The polarization state of the reflected light is determined using the Stokes parameters.
 Such experimental technique was successfully used in our previous studies of other structures~\cite{OptActiv, LVK,Interf}.
 The accuracy of the measured degree of polarization was estimated as 0.1~\%.
The sample holder allowed us to rotate the sample around the normal axis by an angle up to $360^\circ$ for obtaining dependencies of the transmitted light polarization state on the orientation of the incidence plane relative to the crystallographic axes.
For measurements of light transmission, a halogen lamp was used as a light source. A parallel beam of light was formed using lenses and slits. The spectra were registered with a 0.5~m~monochromator and a CCD camera. All transmission measurements were performed at room temperature.  

To visualize the morphology of the samples under study, we used the \textbf{optical microscope} with the possibility of measuring in different polarizations. Surface image with bright-field microscopy obtained by Nikon industrial microscope ECLIPSE LV150. Other details of the optical techniques and results are given in Supplementary Materials.

\section*{Conflicts of interest}
There are no conflicts to declare.

\section*{Acknowledgements}
This work is supported  by the Russian Science Foundation (project No. 19-12-00273). We thank M. Rem\v{s}kar for supplying the synthesized MoS$_2$ flakes and M. A. Yagovkina for probing the samples by X-ray measurements.
The sample characterizations by TEM and polarization microscopy were done at the Federal Joint Research Center “Material science and characterization in advanced technology” supported by the Ministry of Science and Higher Education of the Russian Federation (id RFMEFI62119X0021).




\balance



\providecommand*{\mcitethebibliography}{\thebibliography}
\csname @ifundefined\endcsname{endmcitethebibliography}
{\let\endmcitethebibliography\endthebibliography}{}

\end{document}


\pagestyle{fancy}
\thispagestyle{plain}
\fancypagestyle{plain}{

\fancyhead[C]{\includegraphics[width=18.5cm]{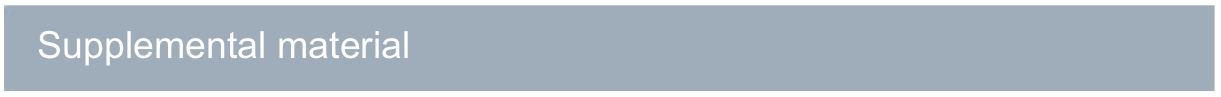}}
\renewcommand{\headrulewidth}{0pt}
}

\makeFNbottom
\makeatletter
\renewcommand\LARGE{\@setfontsize\LARGE{15pt}{17}}
\renewcommand\Large{\@setfontsize\Large{12pt}{14}}
\renewcommand\large{\@setfontsize\large{10pt}{12}}
\renewcommand\footnotesize{\@setfontsize\footnotesize{7pt}{10}}
\makeatother

\renewcommand{\thefootnote}{\fnsymbol{footnote}}
\renewcommand\footnoterule{\vspace*{1pt}%
\color{cream}\hrule width 3.5in height 0.4pt \color{black}\vspace*{5pt}} 
\setcounter{secnumdepth}{5}

\makeatletter 
\renewcommand\@biblabel[1]{#1}            
\renewcommand\@makefntext[1]%
{\noindent\makebox[0pt][r]{\@thefnmark\,}#1}
\makeatother 
\renewcommand{\figurename}{\small{Fig.}~}
\sectionfont{\sffamily\Large}
\subsectionfont{\normalsize}
\subsubsectionfont{\bf}
\setstretch{1.125} 
\setlength{\skip\footins}{0.8cm}
\setlength{\footnotesep}{0.25cm}
\setlength{\jot}{10pt}
\titlespacing*{\section}{0pt}{4pt}{4pt}
\titlespacing*{\subsection}{0pt}{15pt}{1pt}
\pagestyle{plain}
\fancyfoot{}
\fancyfoot[LO,RE]{\vspace{-9.1pt}\includegraphics[height=9pt]{LF}}
\fancyfoot[CO]{\vspace{-9.1pt}\hspace{13.2cm}\includegraphics{RF}}
\fancyfoot[CE]{\vspace{-9.2pt}\hspace{-14.2cm}\includegraphics{RF}}
\fancyfoot[RO]{\footnotesize{\sffamily{1--\pageref{LastPage} ~\textbar  \hspace{2pt}\thepage}}}
\fancyfoot[LE]{\footnotesize{\sffamily{\thepage~\textbar\hspace{3.45cm} 1--\pageref{LastPage}}}}
\fancyhead{}
\renewcommand{\headrulewidth}{0pt} 
\renewcommand{\footrulewidth}{0pt}
\setlength{\arrayrulewidth}{1pt}
\setlength{\columnsep}{6.5mm}
\setlength\bibsep{1pt}

\makeatletter 
\newlength{\figrulesep} 
\setlength{\figrulesep}{0.5\textfloatsep} 

\newcommand{\topfigrule}{\vspace*{-1pt}%
\noindent{\color{cream}\rule[-\figrulesep]{\columnwidth}{1.5pt}} }

\newcommand{\botfigrule}{\vspace*{-2pt}%
\noindent{\color{cream}\rule[\figrulesep]{\columnwidth}{1.5pt}} }

\newcommand{\dblfigrule}{\vspace*{-1pt}%
\noindent{\color{cream}\rule[-\figrulesep]{\textwidth}{1.5pt}} }

\makeatother

\twocolumn[
  \begin{@twocolumnfalse}
\vspace{0.5cm}
\sffamily
\begin{tabular}{m{17cm} p{12.5cm} }

{\textbf{for ``MoS$_2$ flake as a van der Waals homostructure: luminescence properties~and optical anisotropy ''}} \\
\vspace{0.2cm}
 {Lyubov V.  Kotova,$^{\ast}$\textit{$^{a}$} Maxim V. Rakhlin,\textit{$^{a}$} Aidar I. Galimov,\textit{$^{a}$} Ilya A. Eliseyev,\textit{$^{a}$} Bogdan~R.~Borodin,\textit{$^{a}$} Alexey V. Platonov,\textit{$^{a}$} Demid A. Kirilenko,\textit{$^{a}$} Alexander V. Poshakinskiy,\textit{$^{a}$} and Tatiana V. Shubina\textit{$^{a}$}}

\end{tabular}

 \end{@twocolumnfalse} \vspace{0.3cm} ]


\footnotetext{\textit{$^{a}$~Ioffe  Institute, St.~Petersburg, 194021, Russia. E-mail:kotova@mail.ioffe.ru}}

\section{Structure characterization}

\begin{figure*}
\centering
 \includegraphics[width=1.95\columnwidth]{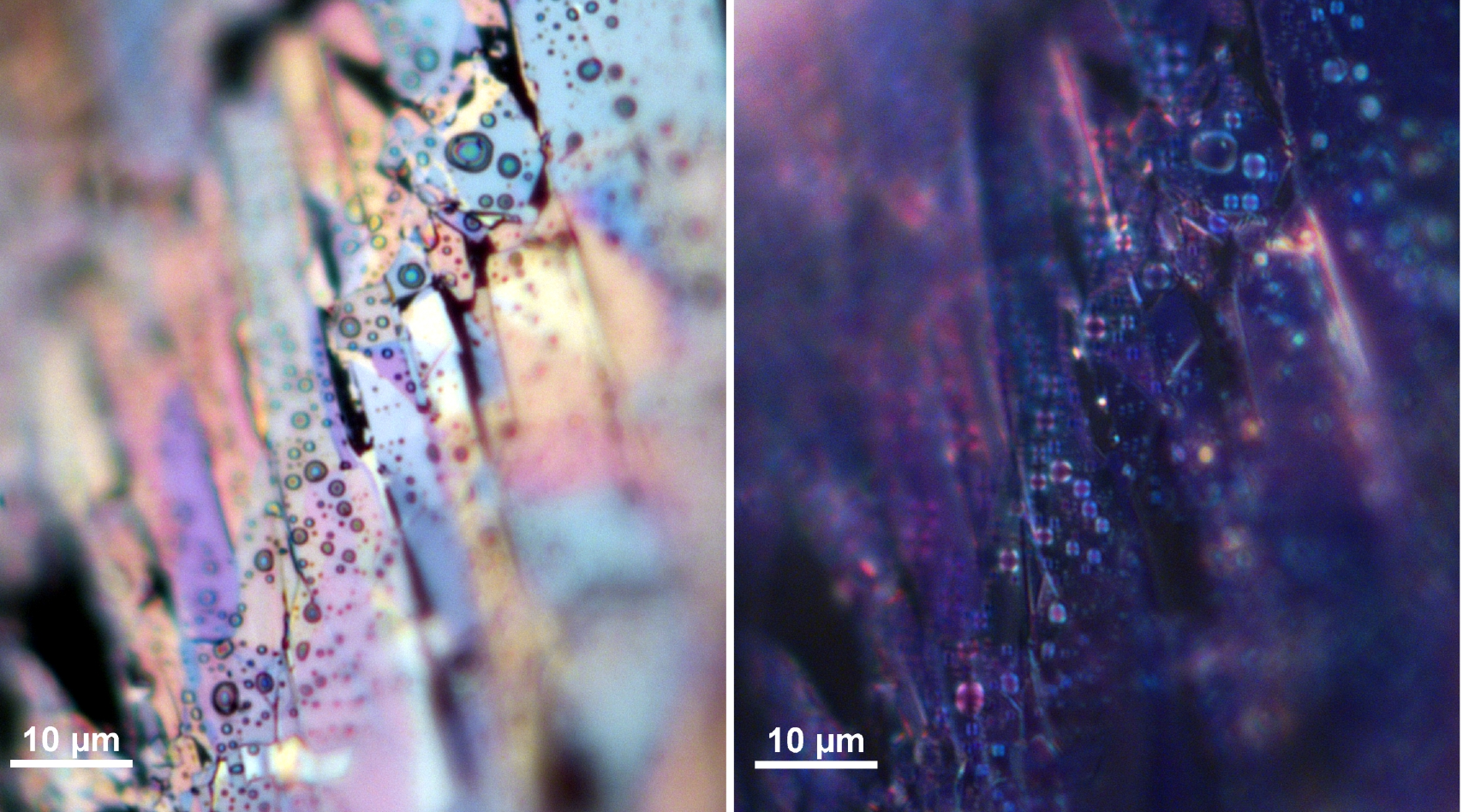}
  \caption{MoS$_2$ flake's surface image with bright/dark field microscopy made by Nikon industrial microscope ECLIPSE LV150.}
  \label{fgr:figura4SUP}
 \end{figure*}

 The studied van der Waals homostructures were fabricated by exfoliation method from bulk MoS$_2$. The surface photograph is shown in Fig.~\ref{fgr:figura4SUP}. The high quality of the bulk crystal was confirmed by TEM diffraction patterns shown in Fig.~\ref{fgr:figura45SUP}.    
 As discussed in the main text, the experimental results indicate that the obtained  structure consist of bulk MoS$_2$ core and few monolayers on the  surface. The monolayers are freely located on the surface of the bulk and joined by van der Waals forces. 
 
 The results of the TEM studies of samples fabricated from a bulk MoS$_2$ crystal by mechanical cleavage are presented in Fig.~\ref{TEM}. We focused on specimens produced from the bulk MoS$_2$ crystal by mechanical cleavage, that is a cleaner method than the adhesive tape exfoliation~\cite{TEM}.The image of a material fragment obtained by cleavage along the (0001) plane is  given in Fig.~\ref{TEM}~a.  A cross-section TEM specimen was made from an adjacent  MoS$_2$ fragment. The high resolution image in Fig.~\ref{TEM}~b exhibits many extended defects. Some of them are the stacking faults and others look as seeds of ripplocations, whose chain is marked by magenta arrows. The ripplocations -- internal ripples -- can appear during cleaving due to the ease of bending of atomic layers and weak van der Waals bonds between them~\cite{Kushima}. In layered crystals, they tend to group together~\cite{Gruber}, promoting the formation of  penetrating hyperdislocations~\cite{Ly}. When the strain of cleaving is high, the continuous layers can be separated into domains by such extended defects.  

 Diffraction patterns of the cross-sectional specimen  Fig.~\ref{TEM}~c,d obtained at different orientations correspond to the lateral (in-plane) displacement of the atomic layers. Moreover, the elongation of the diffraction spots in given patterns increases proportionally to the diffraction index. This indicates the presence of regions with markedly different interlayer spacing, at least 10$\%$  larger than the characteristic value of molybdenum disulfide. The presence of filaments near the spots of type 2$\bar{1}\bar{1}$0 cannot be explained only by a change in the spacing. This assumes the contribution of the lateral displacement of the atomic layers by vectors incommensurate with the lattice period as well.
In general, the TEM results show that sample preparation provides suitable conditions not only for the formation of defects, but also for the separation of atomic layers.  
 
\begin{figure*}
\centering
  \includegraphics[width=1.35\columnwidth]{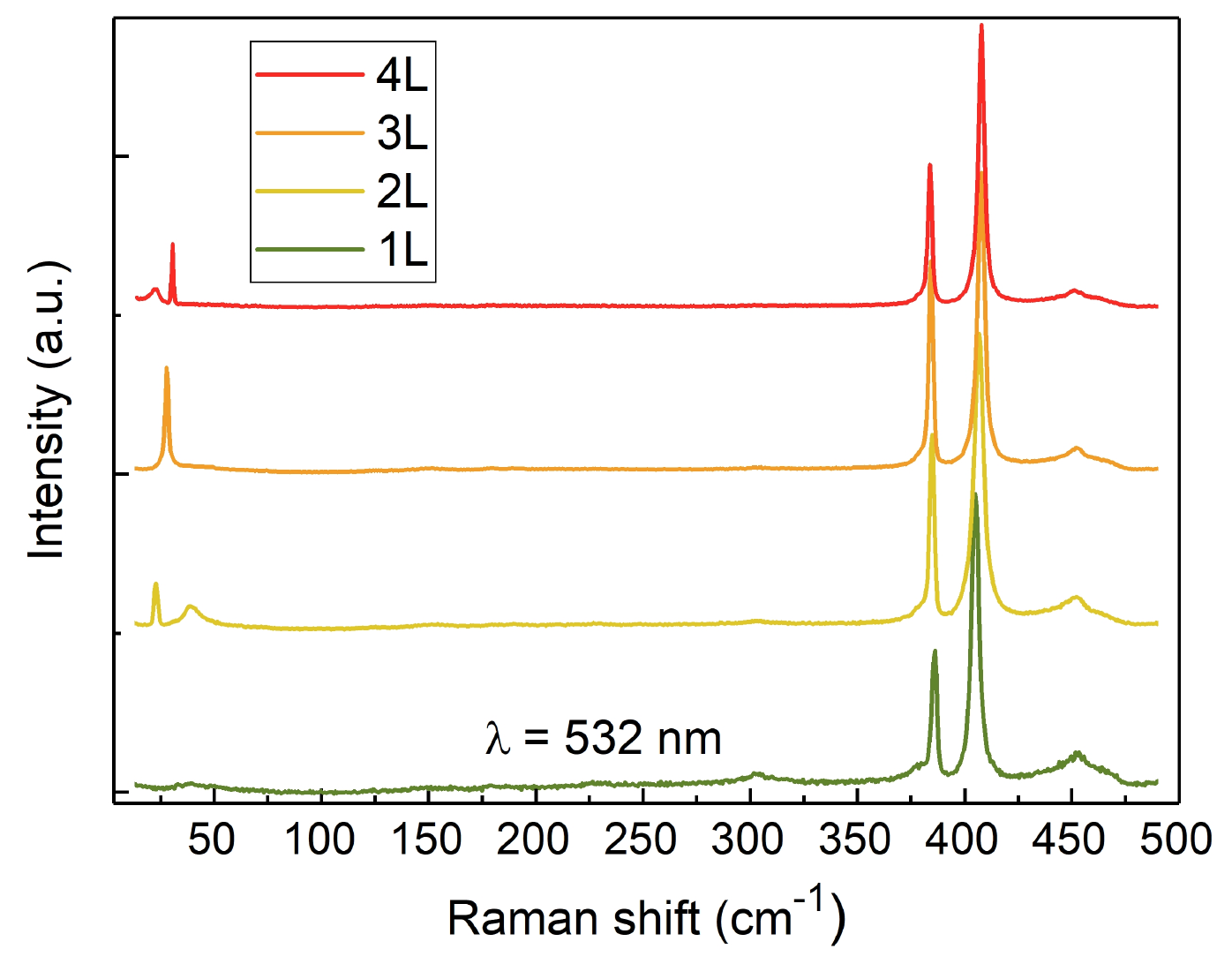}
  \caption{Raman spectra collected from the different point of flake's surface obtained after last step of the exfoliation procedure. Variation of the peak position and intensity is used to determine the number of monolayers on the surface. }
  \label{fgr:figuraSUP_R}
\end{figure*}

\begin{figure*}
\centering
 \includegraphics[width=1.35\columnwidth]{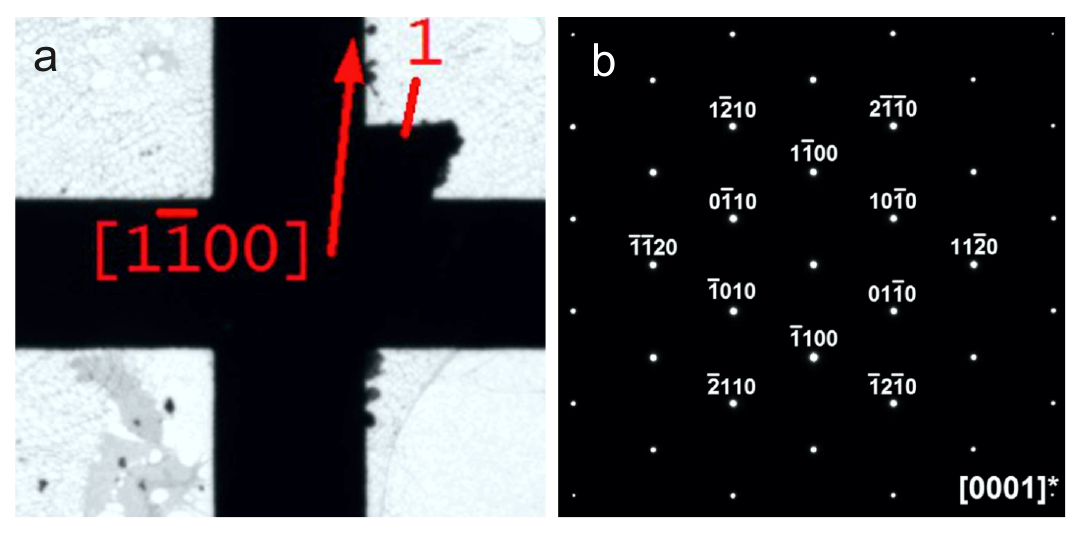}
  \caption{(a) Orientation of the crystallographic axes of the molybdenum disulfide bulk crystal. The arrow indicates the direction [1$\bar{1}$00] determined from the diffraction pattern shown in panel (b). 
  }
  \label{fgr:figura45SUP}
 \end{figure*}

\begin{figure*}
 \centering
  \includegraphics[width=1.35\columnwidth]{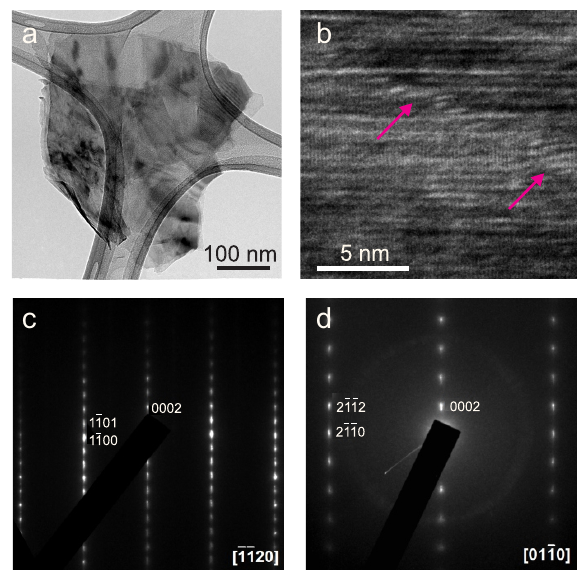}
  \caption{ a) TEM image of a flake cleaved along the (0001) plane. It lies on a carbon lacey film. b) Cross-sectional image of a multilayer specimen cut from the same bulk crystal across the layers. The arising chain of ripples is marked with magenta arrows.  (c), (d) Electron diffraction pattern of a cleaved specimen, the same as in (b), obtained in cross-sectional geometry at various orientations indicated in the graphs. }
  \label{TEM}
\end{figure*}

 Additionally, the AFM images and Raman specrta were measured from the flakes obtained at the last step of the Scotch-tape  procedure. Since the atomic layers of the material are easily peeled off,  a few isolated monolayers appear on the surface after exfoliation. 
 To determine their numbers we have studied the Raman spectra of such detached flakes, see Fig.~\ref{fgr:figuraSUP_R}.  The position and intensity of the C peak in Raman spectra yields that the number of monolayers is varies from 1 to 4 in the studied areas.  AFM studies, shown in main text, give similar results.

\begin{figure*} 
\centering
  \includegraphics[width=1.25\columnwidth]{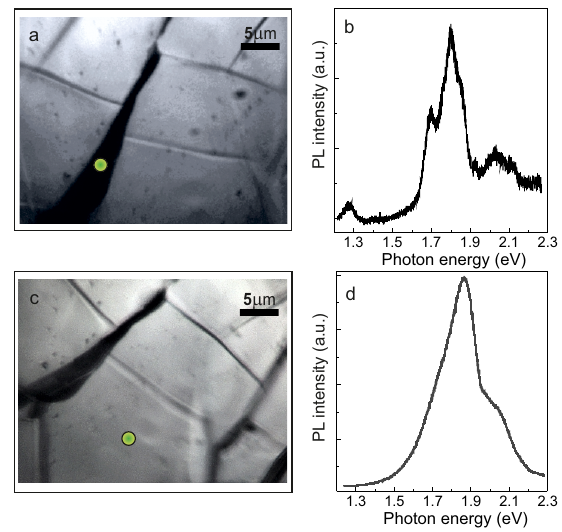}
  \caption{(a,c) MoS$_2$ flake's surface image show the area where micro-PL was excited. (b,d) Micro-PL intensity spectra (T=10K). }
  \label{fgr:figura1SUP}
\end{figure*}

We note that the sample was characterized by micro-photoluminescence experiment in different points of the sample's surface. There are three main types of areas on the sample's surface. The main types are shown and analyzed in the main text. However, there are few special points, which show different photoluminescence spectra, shown in Fig.~\ref{fgr:figura1SUP}.
The first one,  Fig.~\ref{fgr:figura1SUP}~a,b, is detected from the slit and demonstrates features characteristic of bulk MoS$_2$ material. The signal is very weak (as seen from signal-to-noise ratio). In addition to direct transitions, an indirect exciton peak at 1.3\,eV is present.  
Fig.~\ref{fgr:figura1SUP}~d show the PL spectra collected from the surface point shown in Fig.~\ref{fgr:figura1SUP}~c, which is located  away from slits and features 1--2 covering monolayers, resulting in a smaller absorption at exciton resonance as compared to spectra in  Fig. 3(d) and (f) of the main text.

\section{Extra experimental results}

\begin{figure*} 
\centering
  \includegraphics[width=0.9\columnwidth]{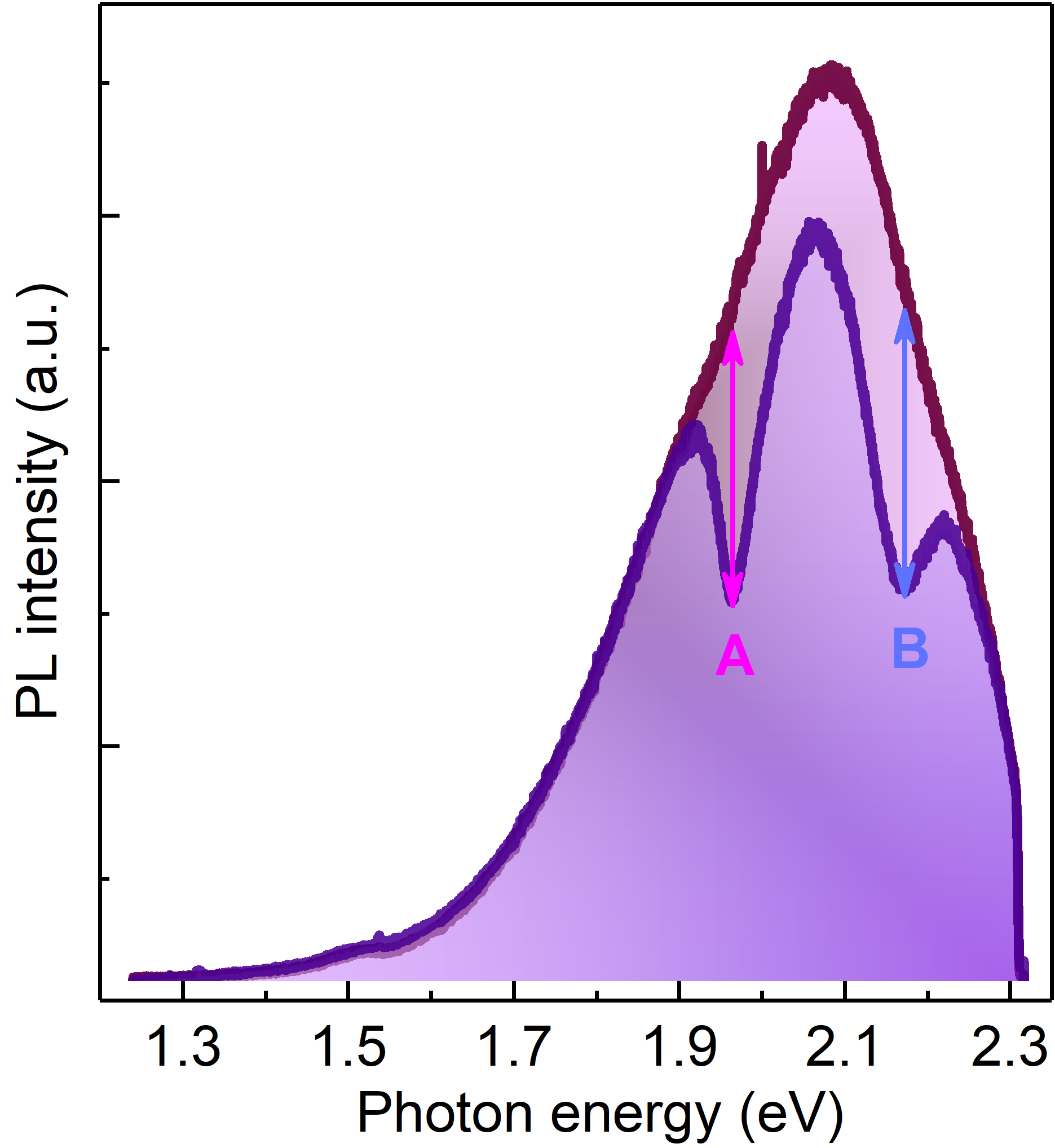}
  \caption{Comparison of the PL spectra collected from surface points with and without monolayers.}
  \label{fgr:figura3SUP2}
\end{figure*}

In the main text, the exciton absorption was estimated from comparison of the PL spectra collected from surface points with and without monolayers. From Fig.~\ref{fgr:figura3SUP2} we get the fraction of absorbed light of 37~\% for A exciton and 45~\% for B exciton. The positions of the exciton features and their temperature dependence was analysed in Fig. 5 (a) of the main text and compared with the  data reported in the literature~\cite{Wilson,Evans,Frindt,Eda,13,frey,Confocal,Mak,nanomat}, that is summarized in Tables~\ref{tbl:T} and \ref{tbl:PL}.

\begin{figure*}
\centering
  \includegraphics[width=1.1\columnwidth]{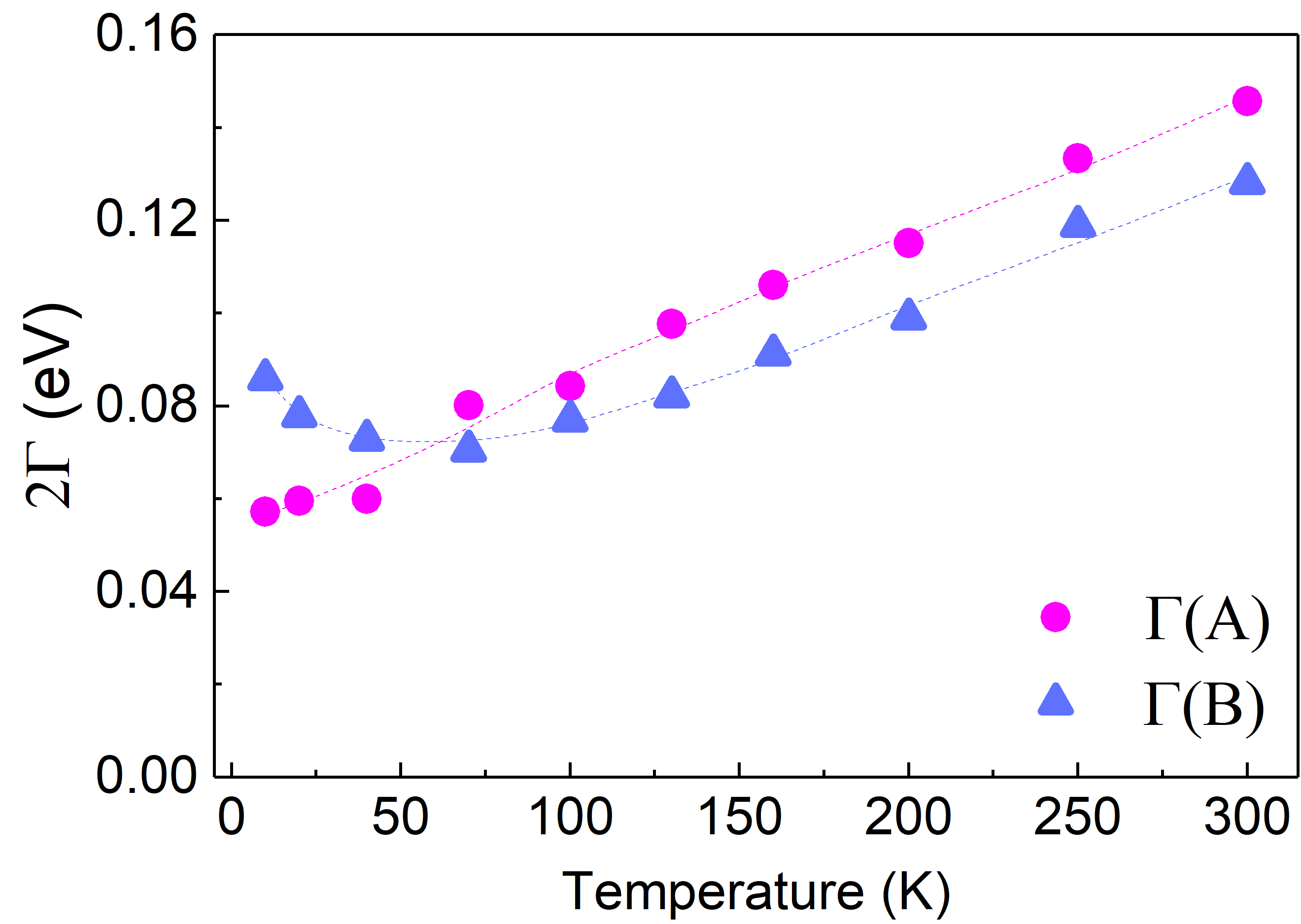}
  \caption{Dependence of the exciton dip width  $\Gamma$ on temperature.}
  \label{fgr:figura3SUP3}
\end{figure*}

Fig.~\ref{fgr:figura3SUP3} show the temperature dependence of the exciton dips FWHM, obtained from the fit of Fig.~4 of the main text.

\begin{figure*}
\centering
  \includegraphics[width=1\columnwidth]{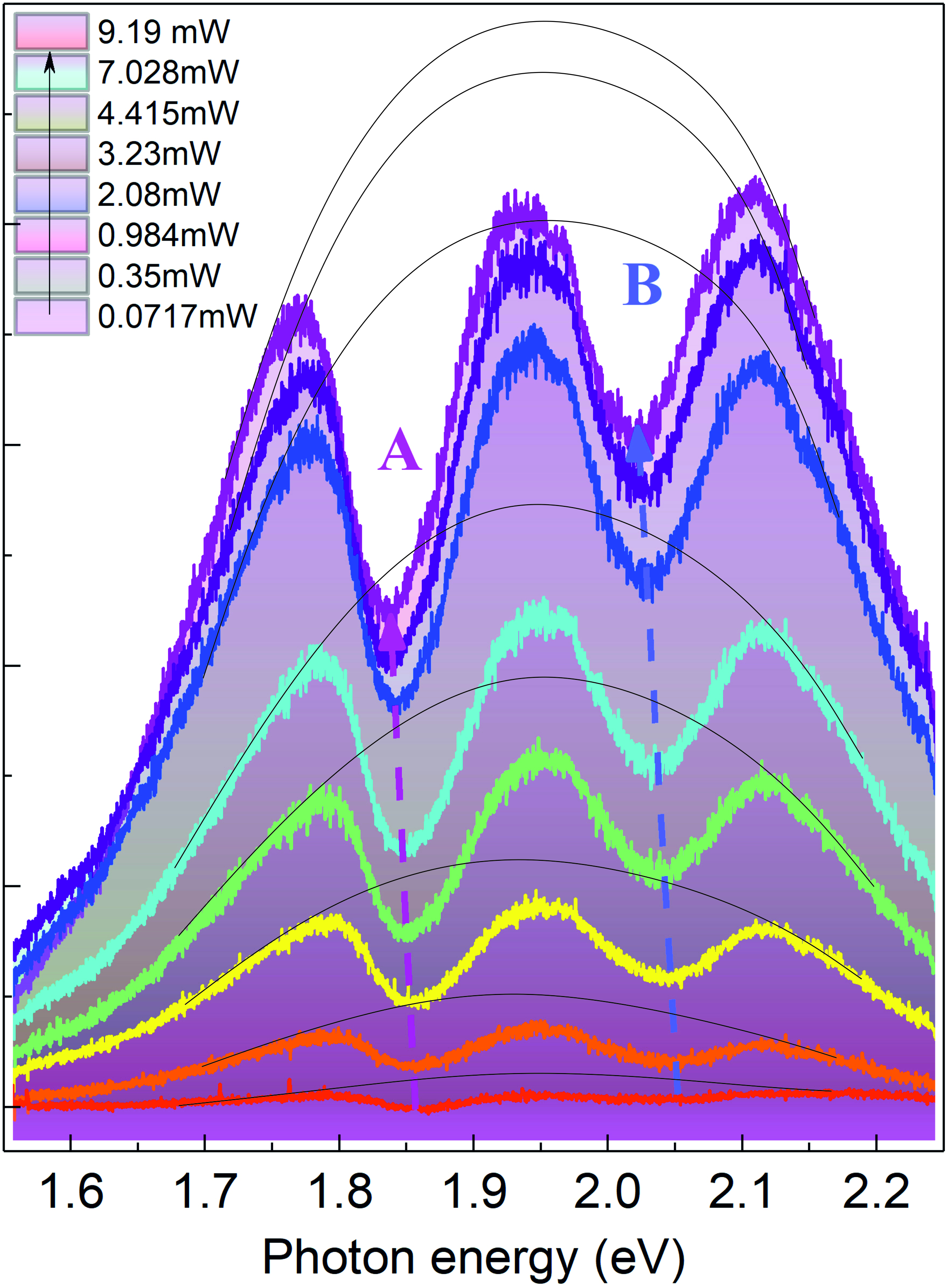}
  \caption{Pump power dependence of the $\mu$-PL spectra obtained at room temperature. The Gaussian envelopes and the  A and B exciton dips are indicated. }
  \label{fgr:figura2SUP}
\end{figure*}

\begin{figure*}
\centering
  \includegraphics[width=1\columnwidth]{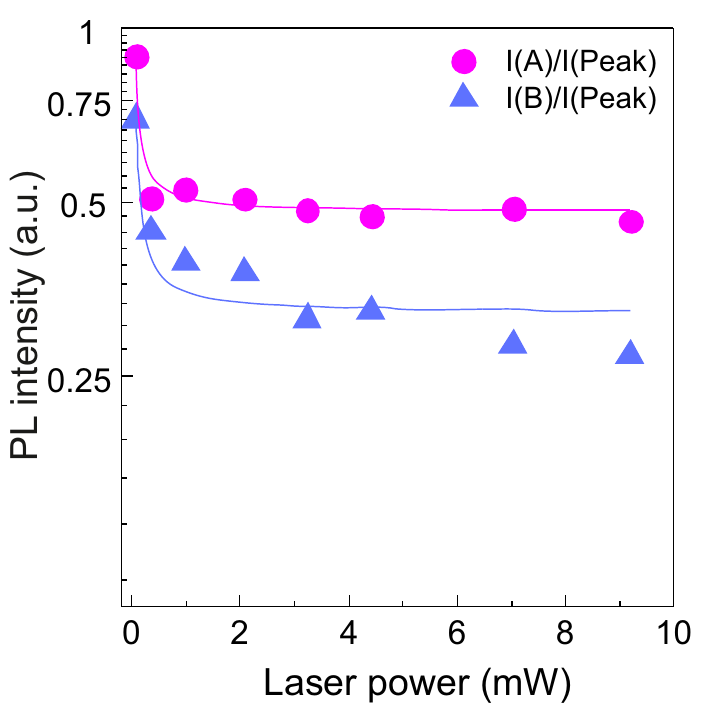}
  \caption{Logarithmic plot of the  exciton dips intensities divided by main PL peak intensity as a function of laser power. }
  \label{fgr:figura3SUP1}
\end{figure*}

 Fig.~\ref{fgr:figura2SUP} illustrates the evolution of the $\mu$-PL spectra with pump power. The broad peak slowly increases with the pump, while the position and relative depth of the dips remain almost constant. The relative depth of the dips is shown in Fig.~\ref{fgr:figura3SUP1}.

\begin{figure*}
\centering
  \includegraphics[width=1.35\columnwidth]{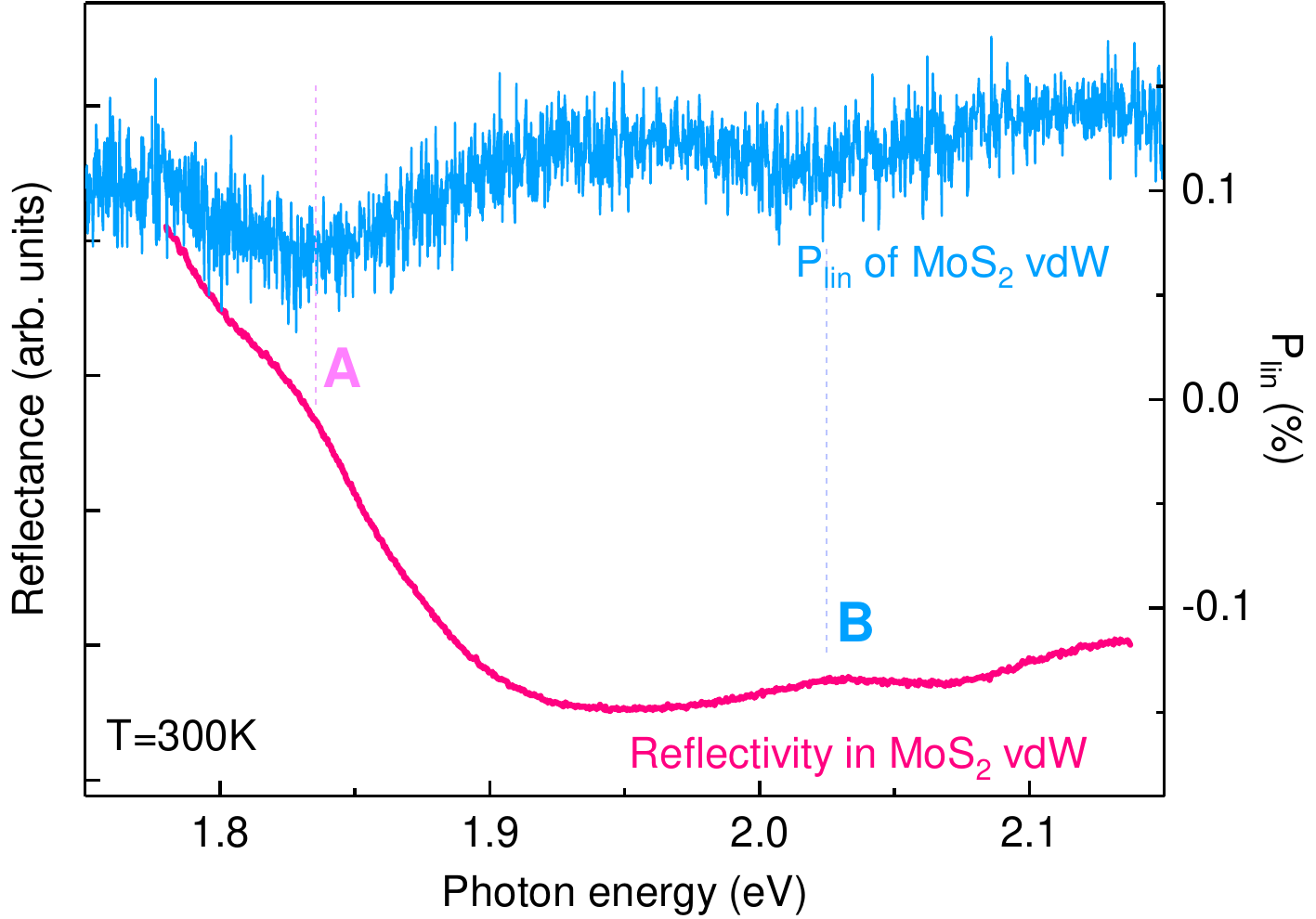}
  \caption{Spectrum of the PL linear polarization degree $P_{\rm lin}$ compared with the reflectance spectrum of the vdW MoS$_2$ structure. Both measurements were done at room temperature. }
  \label{fgr:figura11SUP1}
\end{figure*}

Fig.~\ref{fgr:figura11SUP1} compares the spectrum of the PL linear polarization degree $P_{\rm lin}$ with the reflectance spectrum. Both spectra reveal small peaks at the energies of A and B excitons.

\begin{figure*}
\centering
  \includegraphics[width=1.4\columnwidth]{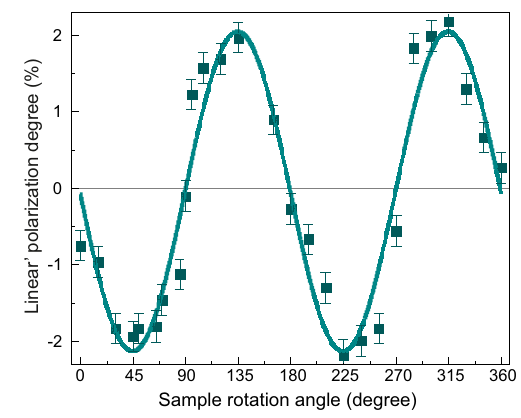}
  \caption{Angular dependence of the transmitted light linear polarization degree in the axes rotated by $\pm 45^\circ$ relative to the incidence plane $\tilde{P}_{\rm lin}$. }
  \label{fgr:figura9SUP}
\end{figure*}

Full Stokes parameters of the transmitted light were measured to characterize optical anisotropy of the vdW homostructure. Two of them are presented in the main text, and the remaining, $\tilde{P}_{\rm lin}$ corresponding to the linear polarization degree in the axes rotated by $\pm 45^\circ$ relative to the incidence plane, is shown in Fig.~\ref{fgr:figura9SUP}. The polarization degree is small (less 2~\%) as compared to background contribution to ${P}_{\rm circ}$, indicating the absence of the linear dichroism.


\providecommand*{\mcitethebibliography}{\thebibliography}
\csname @ifundefined\endcsname{endmcitethebibliography}
{\let\endmcitethebibliography\endthebibliography}{}

\begin{table*}
\small
  \caption{\ Photoluminescence. \textendash\ Exciton's position.}
  \label{tbl:PL}
  \begin{tabular*}{\textwidth}{@{\extracolsep{\fill}}lllllll}
    \hline
    Sample & Temperature & A exciton & B exciton & Reference \\
    \hline
     MoS$_2$ thin (1.3 nm) films  & 300 K & 1.878 eV  & 2.03 eV   &~\cite{Eda}\\
    Bulk & 10 K& 1.82 eV  & 2.01 eV   &~\cite{Confocal} \\
    Trilayer & 10 K & 1.83 eV  &  2.01 eV   &  \\
    Bilayer &10 K & 1.86 eV  & 2.02 eV  &  \\
       Monolayer & 10 K& 1.88 eV  & 2.03 eV  &  \\
     Monolayer &300 K & 1.82 eV  & 2.01 eV   &~\cite{Mak} \\
      Bilayer &300 K & 1.86 eV  & 2.02 eV  &  \\
       Monolayer  MoS$_2$ on the SiO$_2$/Si substrate & 300 K &  1.831 eV &  1.977 eV &  ~\cite{nanomat}\\
      
    \hline
  \end{tabular*}
\end{table*}

\begin{table*}
\small
  \caption{\ Transmission. \textendash\ Exciton's position}
  \label{tbl:T}
  \begin{tabular*}{\textwidth}{@{\extracolsep{\fill}}lllllll}
    \hline
    Sample & Temperature & A exciton & B exciton &  Reference \\
    \hline
    MoS$_2$ thin crystals &  77 K & 1.921eV  & 2.02 eV  & ~\cite{Wilson} \\
     & 293 K & 1.86 eV  &2.1076 eV  &  \\
    0.06 $\mu$m flake & 77 K &   1.971 eV  &  2.163 eV   &~\cite{Evans}  \\
         & 293 K & 1.88 eV  & 2.072 eV  &   \\
    MoS$_2$ thin crystals & 77 K & 1.92 eV  & 2.122 eV   &~\cite{Frindt}\\
      & 293 K &   1.861 eV  & 2.048 eV  & \\
MoS$_2$ thin (1.3 nm) films  & 300 K & 1.878 eV  & 2.03 eV   &~\cite{Eda}\\
2H-MoS$_2$  & 300 K & 1.88 eV  & 2.06 eV   &~\cite{13}\\
2H-MoS$_2$  & 150 K & 1.90 eV  & 2.10 eV   &~\cite{frey}\\
2H-MoS$_2$  & 75 K & 1.91 eV  & 2.11 eV   & \\
2H-MoS$_2$  & 25 K & 1.91 eV  & 2.11 eV   & \\

       MoS$_2$ van der Waals homostructures & 297 K & 1.86 eV  & 2.01 eV  &  our data in transmission\\
       MoS$_2$ van der Waals homostructures & 297 K &  1.83 eV  & 2.016 eV  &  our data in PL \\
       MoS$_2$ van der Waals homostructures & 77 K &  1.929 eV  & 2.137 eV  &  our data in PL \\
       MoS$_2$ van der Waals homostructures & 10 K &  1.94 eV  & 2.16 eV  &  our data in PL \\

    \hline
  \end{tabular*}
\end{table*}